\documentclass[prd,preprint,nofootinbib]{revtex4}
\usepackage{graphicx, cancel}

\begin{document}

\newcommand{\gsim}{ \mathop{}_{\textstyle \sim}^{\textstyle >} }
\newcommand{\lsim}{ \mathop{}_{\textstyle \sim}^{\textstyle <} }
\newcommand{\vev}[1]{ \left\langle {#1} \right\rangle }

\newcommand{\bear}{\begin{array}}  \newcommand{\eear}{\end{array}}
\newcommand{\bea}{\begin{eqnarray}}  \newcommand{\eea}{\end{eqnarray}}
\newcommand{\beq}{\begin{equation}}  \newcommand{\eeq}{\end{equation}}
\newcommand{\bef}{\begin{figure}}  \newcommand{\eef}{\end{figure}}
\newcommand{\bec}{\begin{center}}  \newcommand{\eec}{\end{center}}
\newcommand{\non}{\nonumber}  \newcommand{\eqn}[1]{\beq {#1}\eeq}
\newcommand{\la}{\left\langle} \newcommand{\ra}{\right\rangle}
\newcommand{\ds}{\displaystyle}

\def\SEC#1{Sec.~\ref{#1}}
\def\FIG#1{Fig.~\ref{#1}}
\def\EQ#1{Eq.~(\ref{#1})}
\def\EQS#1{Eqs.~(\ref{#1})}
\def\lrf#1#2{ \left(\frac{#1}{#2}\right)}
\def\lrfp#1#2#3{ \left(\frac{#1}{#2}\right)^{#3}}
\def\GEV#1{10^{#1}{\rm\,GeV}}
\def\MEV#1{10^{#1}{\rm\,MeV}}
\def\KEV#1{10^{#1}{\rm\,keV}}

\def\lrf#1#2{ \left(\frac{#1}{#2}\right)}
\def\lrfp#1#2#3{ \left(\frac{#1}{#2}\right)^{#3}}

%

\renewcommand{\thefootnote}{\alph{footnote}}

\renewcommand{\thefootnote}{\fnsymbol{footnote}}
\preprint{DESY 07-061}
\title{Inflaton Decay in Supergravity}
\renewcommand{\thefootnote}{\alph{footnote}}

\author{Motoi Endo$^{1}$, Fuminobu Takahashi$^{1}$, and T. T. Yanagida$^{2,3}$}

\affiliation{
${}^1$ Deutsches Elektronen Synchrotron DESY, Notkestrasse 85,
  22607 Hamburg, Germany\\
${}^2${\it Department of Physics, University of Tokyo,
     Tokyo 113-0033, Japan}\\
${}^3${\it Research Center for the Early Universe, University of Tokyo,
     Tokyo 113-0033, Japan}
  }

\begin{abstract}
\noindent
We discuss inflaton decay in supergravity, taking account of the
gravitational effects. It is shown that, if the inflaton has a nonzero
vacuum expectation value, it generically couples to any matter fields
that appear in the superpotential at the tree level, and to any gauge
sectors through anomalies in the supergravity. Through these
processes, the inflaton generically decays into the supersymmetry
breaking sector, producing many gravitinos. The inflaton also directly
decays into a pair of the gravitinos. We derive constraints on both
inflation models and supersymmetry breaking scenarios for avoiding 
overproduction of the gravitinos. Furthermore, the inflaton naturally 
decays into the visible sector via the top Yukawa coupling and 
SU(3)$_{C}$ gauge interactions. 
\end{abstract}

\pacs{98.80.Cq,11.30.Pb,04.65.+e}

\maketitle

\section{Introduction}
\label{sec:1}

Inflation~\cite{Guth:1980zm} provides a simple solution to a number of
serious shortcomings in the big bang cosmology such as the horizon and
flatness problems. Above all, it can account for the origin of density
fluctuations necessary to form the rich structure of our universe. In
fact, the standard slow-roll inflation predicts almost scale-invariant
power spectrum, which fits the recent cosmic microwave background
(CMB) data~\cite{Spergel:2006hy} quite well.

It is now recognized that the universe underwent an inflationary epoch
at an early stage. During the inflation, the universe is dominated by
the potential energy of the inflaton, and experiences exponential
expansion~\cite{Sato:1980yn,Guth:1980zm}. After inflation ends, the
inflaton field releases its energy into a thermal plasma by the decay,
and the universe is reheated. Since all the particles including photons
and baryons in the present universe are ultimately originated from the
inflaton decay, it is of great importance to reveal how the reheating
proceeds.

So far however, the reheating process has not been fully investigated.
One often simplifies the whole reheating processes, and expresses them
in terms of a single parameter, the reheating temperature.  That is,
the inflaton is assumed to have some ad hoc interactions with lighter
degrees of freedom, i.e., the standard model (SM) particles in most
cases, while possible productions of the hidden fields and/or
gravitinos are neglected without definite grounds.  However, many
cosmological phenomena, e.g., baryogenesis, and production of dark
matter and unwanted relics, crucially depend on the details of the
reheating. Although the reheating temperature is certainly an
important characteristic parameter, such simplification is too crude
to truly describe cosmological scenarios.

Recently there has been much progress concerning the decays of scalar
fields such as moduli~\cite{moduli,Dine:2006ii,Endo:2006tf} and
inflaton~\cite{Kawasaki:2006gs,Asaka:2006bv,Endo:2006qk,Endo:2006xg,Endo:2007ih}
in a framework of the local supersymmetry (SUSY), i.e., the
supergravity (SUGRA). The supersymmetric extension is one of the most
promising candidates for the theory beyond SM. If SUSY exists at the
TeV scale, the inflaton dynamics is quite likely described in SUGRA.
In addition, since the existence of a flat direction is mediocre in
SUSY models, one can find extremely flat potentials appropriate for
the slow-roll inflation.  Throughout this paper we consider inflation
models in SUGRA. We have investigated the reheating of the universe 
in this framework, and found that the gravitinos are generically 
produced from the inflaton decay in most inflation models. In particular,
Ref.~\cite{Kawasaki:2006gs} has first pointed out that the inflaton
can directly decay into a pair of the gravitinos. Moreover,
incorporating the gravitational effects,
Refs.~\cite{Endo:2006qk,Endo:2007ih} have shown that the inflaton
generically decays into the SUSY breaking sector, which produces the
gravitinos (in)directly. The gravitino production rates due to these
processes depend on the inflaton parameters as well as the
detailed structure of the SUSY breaking sector. Such gravitino
production clearly goes beyond the simplification of the reheating
that has been adopted so far, and interestingly enough, it provides
severe constraints on inflation models as well as the SUSY breaking
scenarios. These constraints, together with the future collider
experiments and observations on CMB, should become an important 
guide to understand the high energy physics and the early universe. 
The purpose of the present paper is to provide a global picture  
of the inflaton decay processes in SUGRA, paying special attention 
to the gravitino production. In particular, we explain which decay
processes become most important under which circumstances. 
Not only do we summarize the decay processes found so far 
but we also give complete results on the spontaneous decay
and the anomaly-induced decay processes,
including the higher dimensional terms in the K\"ahler
potential and the K\"ahler and sigma-model anomalies.

The organization of the paper is as follows. In Sec.~\ref{sec:2} we
review the gravitino pair production at the inflaton decay. Then we
discuss the spontaneous decay at the tree level in Sec.~\ref{sec:3}.
In Sec.~\ref{sec:4} we consider the anomaly-induced decay of the
inflaton, which proceeds via the anomalies in SUGRA. We provide some
results on the decay rates, by way of illustration, for the minimal
and sequestered K\"ahler potentials there.  In Sec.~\ref{sec:5}, we
study cosmological implications of the decay processes explained in
the preceding sections, particularly focusing on the constraints on
the inflation models.  The last section is devoted to conclusion.

\section{Decay into a pair of gravitinos}
\label{sec:2}

Once the inflaton field obtains a finite vacuum expectation value
(VEV), it necessarily decays into the gravitinos. In this section, we
briefly discuss the production of a pair of the gravitinos, and
provide the partial decay rate.  The process we consider is a
perturbative decay, and the gravitinos are produced directly from the
inflaton.  The gravitino production is represented by the following
interactions in the SUGRA Lagrangian~\cite{WessBagger}~\footnote{ Due
to the K\"ahler invariance, the generalized K\"ahler potential $G$ is
more convenient and transparent than using the K\"ahler potential $K$
and the superpotential $W$. Since these two frames are related by the
Weyl transformation, any physical amplitudes are equivalent at the
tree level.  };
\begin{eqnarray}
    e^{-1}\mathcal{L} &=&
    \frac{1}{4} \epsilon^{k\ell mn}
    \left( G_i \partial_k \phi^i - 
      G_{i^*} \partial_k \phi^{*i} \right)
   \bar\psi_\ell \bar\sigma_m \psi_n
   \nonumber\\
   &&
   - \frac{1}{2} e^{G/2} 
   \left( G_i \phi^i + G_{i^*} \phi^{*i} \right)
   \left[ \psi_m \sigma^{mn} \psi_n + 
     \bar\psi_m \bar\sigma^{mn} \bar\psi_n 
   \right],
\end{eqnarray}
where $\sigma^{mn} = \frac{1}{4} (\sigma^m \bar\sigma^n - \sigma^n
\bar\sigma^m)$, and we have chosen the unitary gauge in the Einstein
frame. The sum over the indices is understood unless otherwise stated.
We have also adopted the Planck unit $M_P = 1$ ($M_P = 2.4 \times
10^{18}$GeV) here and in what follows unless it is written explicitly.
The 2-spinor, $\psi_m$ (or $\psi_{3/2}$), represents the gravitino,
while $\phi^i$ collectively denotes an arbitrary scalar field
including the inflaton $\phi$.  Then the decay rate of the inflaton
into a pair of the gravitinos, $\phi \to 2 \psi_{3/2}$, is evaluated
as~\cite{moduli}
\begin{eqnarray}
   \Gamma^{({\rm grav})} &\simeq&
   \frac{|G_\phi|^2}{288\pi} 
   \frac{m_\phi^5}{m_{3/2}^2M_P^2},
\end{eqnarray}
where $m_{3/2} = e^{G/2}$ and $m_\phi$ are the masses of the gravitino
and the inflaton, respectively. We readily find that the decay
amplitude is inversely proportional to $m_{3/2}$.  This is a result of
an enhancement ($\propto m_{3/2}^{-2}$) due to the longitudinal mode
of the gravitino, $\psi_m(k) \propto k_m/m_{3/2} \sim m_\phi/m_{3/2}$,
which is partially compensated by the chirality suppression of the
amplitude ($\propto m_{3/2}$).

The decay amplitude crucially depends on $G_\phi$, which is a
derivative of the generalized K\"ahler potential, $G = K + \ln |W|^2$,
with respect to the inflaton field $\phi$. It is related to an
$F$-term of the inflaton supermultiplet through the equation of
motion, $F^i = - e^{G/2} g^{ij^*} G_{j^*}$. In order to evaluate
$G_\phi$, we need to incorporate the SUSY breaking field, $z$, into
our analysis.  This is because of the following reason. The decay is
treated in the mass-eigenstate basis. In this basis, $\phi$ generally
mixes with $z$ due to the SUGRA effects, unless the inflaton is
protected by some symmetries which are preserved at the vacuum.  We
take $G_z = O(1)$ to have the vanishing cosmological constant. Then,
$G_z$ can contribute to $G_\phi$ effectively via mixings between
$\phi$ and $z$, which enhances gravitino production rate from inflaton
decay. That is, the inflaton first oscillates into $z^{(*)}$, which
then decays into a pair of the gravitinos: $\phi \rightleftharpoons
z^{(*)} \to 2 \psi_{3/2}$.

The mixing angle depends on the mass spectrum of $\phi$ and $z$. The
direct pair-gravitino production is effective especially for $m_\phi
\ll m_z$. Such a large $m_z$ is often realized in the dynamical SUSY
breaking (DSB) scenario~\cite{Witten:1981nf}. In this case, there is a
soft mass term, $K \sim |z|^4 / \Lambda^2$ ($\Lambda$ is the dynamical
scale), and a scalar mass of $z$ can be larger than $m_\phi$,
depending on inflation models. Then $G_\phi$ is given
by~\cite{Endo:2006tf}
\begin{eqnarray}
    \left| G_\phi \right|^2 &\simeq& 
    \left| \sqrt{3} g_{\phi z^*} \right|^2 + 
    \left| \sqrt{3} (\nabla_\phi G_z) 
      {m_{3/2} \over m_\phi} \right|^2,
\label{eq:gphi}      
\end{eqnarray}
where we have neglected interference terms and higher dimensional
operators in the K\"ahler potential.  Here $g_{ij^*} =
\frac{\partial^2 K}{\partial \phi^i \partial \phi^{*j}}$ and $\nabla_i
G_j = G_{ij} - \Gamma_{ij}^k G_k$ with $\Gamma^i_{jk} = g^{i\ell^*}
g_{j\ell^* k}$. Note that the first term is from the mixing in the
kinetic terms, while the SUGRA effects contribute to the second
one. Thus even if there is no direct coupling between the inflaton and
SUSY-breaking sectors in the global SUSY limit, the inflaton decays
into a pair of the gravitinos for $m_\phi \ll m_z$.

Since each term of (\ref{eq:gphi}) is expected to depend 
on $\phi$ linearly, it is convenient to express the mixings as
\bea
|\nabla_\phi G_z| &\equiv& c\, \langle \phi \rangle,\non\\
|g_{\phi z^*}| &\equiv& \tilde{c} \,  \langle \phi \rangle.
\label{eq:parameterize}
\eea
In SUGRA, $c$ is estimated to be $O(1)$ for a generic K\"ahler
potential by using $G_z = O(1)$, while $\tilde{c}$ depends on details
of the SUSY breaking sector such as the VEV $\la z \ra$, e.g. ${\tilde
c} = \la z \ra$ for $\delta K = |\phi|^2 |z|^2$.  Then, if $\tilde{c}$
is suppressed as in case of the minimal K\"ahler potential
(i.e. $g_{\phi z^*} = 0$), the gravitino pair production rate is
\begin{eqnarray}
   \Gamma^{({\rm grav})} &\simeq&
   \frac{c^2}{96\pi} 
   \left( \frac{\langle \phi \rangle}{M_P} \right)^2
   \frac{m_\phi^3}{M_P^2}.
   \label{eq:pair-grav}
\end{eqnarray}
On the other hand, if the kinetic mixing is large, the rate is much
enhanced as
\beq
  \Gamma^{({\rm grav})} \;\simeq\; \frac{\tilde{c}^2}{96\pi} 
   \left( \frac{\langle \phi \rangle}{M_P} \right)^2
   \frac{m_\phi^5}{m_{3/2}^2 M_P^2}.
   \label{eq:pair-grav2}
\eeq
Such large gravitino production rates are cosmologically disastrous,
which will be discussed in Sec.~\ref{sec:4}.

For high-scale inflation models with $m_\phi \gg m_z$, the
pair-gravitino production rate depends on the detailed structure of
the SUSY breaking models. If the SUSY breaking field is singlet and
elementary above $\Lambda$~\footnote{
  Such a singlet SUSY breaking field is necessary for the gauginos to
  have a sizable mass in the gravity-mediated SUSY breaking
  scenario~\cite{ModuliProblem,Banks:1993en}.  See also
  Ref.~\cite{Endo:2007cu} for a retrofitted gravity-mediation.
  \label{footnote:gravity-med}
}~\footnote{ 
  In the DSB scenarios, the Polonyi problem was once solved since the
  $z$ field decays much before BBN due to a large soft scalar mass of
  $z$~\cite{ModuliProblem,Banks:1993en}.  However, it has been
  recently found that the presence of such a field still puts a severe
  bound on the inflation scale~\cite{Ibe:2006am}.
  \label{footnote:Polonyi}
}, the inflaton still directly decays into a pair of the
gravitinos. In this case the relevant contribution to $G_\phi$ comes
from higher dimensional terms, $K \sim (\kappa/2) |\phi|^2 zz + {\rm
h.c.}$~\footnote{
  The contribution from this operator is suppressed when $m_\phi$ is
  smaller than $m_z$~\cite{Endo:2006tf}.
}.  Then the gravitino production rate is given by
(\ref{eq:pair-grav}) with $c$ replaced with $\kappa$ (see
Ref.~\cite{Kawasaki:2006gs,Endo:2006tf} for details).  On the other
hand, if the SUSY breaking field is composed of other fields and if
the dynamical scale $\Lambda$ is below $m_\phi$, the direct production
of the gravitinos becomes suppressed. Instead, as discussed in the
following sections, gravitational effects force the inflaton to decay
into the SUSY breaking sector.

Finally let us make a comment. In addition to the pair-gravitino
production, the gravitino may be singly produced at the decay.  This
is the case when the inflaton mass comes from the soft SUSY breaking
terms. Then the rate becomes as large as that given by
(\ref{eq:pair-grav2}) with $\tilde{c} = O(1)$. However, if the
inflaton mass is provided by a SUSY-invariant mass term (as in most
inflation models), such a single-gravitino production is negligible.

\section{Spontaneous Decay}
\label{sec:3}

In this section we review the spontaneous decay of the inflaton,
$\phi$, at the tree level. If the reheating is induced by the inflaton
decay through non-renormalizable interactions, the reheating
temperature can be low enough to satisfy the constraints from
gravitinos produced by thermal
scatterings~\cite{Weinberg:zq,Krauss:1983ik}. Since the interactions
are then quite weak, the SUGRA effects may play an important
role. Indeed, it has been recently pointed out that the SUGRA effects
induce the inflaton decay~\cite{Endo:2006qk}. The relevant channels of
the inflaton decay contains the 2- and 3-body final states~\footnote{
Although $\phi$ may decay into 4-scalars, it is suppressed by the
phase space and can be neglected.  }.

For the matter-fermion production, the relevant interactions are
provided in the Einstein frame as~\cite{WessBagger}
\begin{eqnarray}
    e^{-1} \mathcal{L} &=& 
    -i g_{ij^*} \bar\chi^j \bar\sigma^\mu \partial_\mu \chi^i
    \nonumber \\ && 
    + \frac{1}{4} g_{ij^*} i(K_k \partial_\mu \phi^k 
      - K_{k^*} \partial_\mu \phi^{*k}) 
      \bar\chi^j \bar\sigma^\mu \chi^i
    -i g_{ij^*} \Gamma^i_{k\ell} (\partial_\mu \phi^k) 
      \bar\chi^j \bar\sigma^\mu \chi^\ell
    \nonumber \\ && 
    -\frac{1}{2} e^{K/2} (\mathcal{D}_i D_j W) \chi^i \chi^j 
    + {\rm h.c.},
    \label{eq:Lagrangian}
\end{eqnarray}
where $\mathcal{D}_i D_j W = W_{ij} + K_{ij} W + K_i D_j W + 
K_j D_i W - K_i K_j W - \Gamma^k_{ij} D_k W$ with $D_i W = W_i 
+ K_i W$. On the other hand, the matter-scalar production is 
represented by the kinetic term and the scalar potential;
\begin{eqnarray}
    e^{-1} \mathcal{L} &=&  -g_{ij^*}
\partial_\mu \phi^i \partial^\mu \phi^{*j} \non\\
&&    - e^{K} \left[ g^{ij^*} (D_i W) (D_j W)^* - 3 |W|^2 \right].
\end{eqnarray}
In this paper, we adopt notation that, when the inflaton $\phi$ is
explicitly shown in expressions, $\phi^i$ and $\chi^i$ represent only
the matter fields. (We also use $Q$ to denote the matter fields.)
Otherwise, as mentioned before, $\phi^i$ collectively denotes an
arbitrary field including the inflaton $\phi$.

First let us consider decay processes induced by higher dimensional
operators. The relevant one arises from such terms in the K\"ahler
potential that holomorphically depend on the matter fields, $Q, Q'$,
i.e.  $\delta K \sim |\phi|^2 QQ' + {\rm h.c.}$~\footnote{
 The decay process from this operator is obtained also in the global
 SUSY models.
}.  The presence of such operators strongly depends on symmetries of
the visible/hidden sectors. In Eq.~(\ref{eq:Lagrangian}) the term
including $\Gamma^\phi_{ij}$ in the fermion mass is given by
\begin{eqnarray}
    \mathcal{L} &=& 
    \frac{1}{2} e^{K/2} 
    g_{\phi^* ij} g^{\phi\phi^*} W_{\phi\phi}\,
    \phi \chi^i \chi^j + {\rm h.c.},    
\label{hol-fermion}    
\end{eqnarray}
which induces the inflaton decay into the two fermions ($\phi \to
\bar\chi^i\bar\chi^j$).  Note here that $e^{K/2} g^{\phi\phi^*}
W_{\phi\phi} \equiv m_\phi$ is the inflaton mass.

On the other hand, the decay into the two scalars ($\phi \to
\phi^i\phi^j$) arises from the kinetic term of the matter scalars,
\begin{eqnarray}
    \mathcal{L} &=& 
    \frac{1}{2} g_{\phi i^*j^*} 
    (\partial^2 \phi) \phi^{*i} \phi^{*j} + {\rm h.c.}.
\label{hol-scalar}    
\end{eqnarray}
Using the equation of motion, $\partial^2 \phi = m_\phi^2 \phi$, one
finds that the decay rates satisfy $\Gamma(\phi \to
\phi^i\phi^j) \simeq \Gamma(\phi \to \bar\chi^i\bar\chi^j)$. The
total rate then becomes
\begin{eqnarray}
    \Gamma^{({\rm 2-body;hol})} &\equiv&
    \Gamma(\phi \to  \bar\chi^i\bar\chi^j) + 
    \Gamma(\phi \to \phi^i\phi^j)
    \non \\
    &\simeq&
    \frac{|g_{\phi^* ij}|^2 }{8\pi} m_{\phi} 
    \left( 1 - {4 M_Q^2 \over m_\phi^2} \right)^{\frac{1}{2}},
    \label{eq:2-body-hol}
\end{eqnarray}
where $M_Q$ is a mass of the final state particles. Here note that $i$
and $j$ are fixed and the sum is not taken over these variables in the
last expression.

Next let us discuss the case of $\Gamma^\phi_{ij} = 0$, which is due
to some symmetries imposed on the $i$- and $j$-matter fields. The
2-body decay then becomes suppressed by the mass of the final-state
particles, $M_Q$. From the Lagrangian (\ref{eq:Lagrangian}), the
effective interaction is given by
\begin{eqnarray}
    \mathcal{L} &=& 
    - \frac{1}{2} e^{K/2} 
    \left(
      K_\phi W_{ij} + W_{\phi ij}
      - 2 \Gamma^k_{\phi i} W_{jk}
    \right)\, \phi \chi^i \chi^j + {\rm h.c.},
    \label{eq:int-1}
\end{eqnarray}
where we have assumed that $|D_\phi W| \ll |W|$. Here and in the
followings, we assume that the matter fields are charged under some
symmetries for simplicity. Then we can set $K_i, W_i \ll 1$ for the
matter fields.  It should be noticed that the second term in the
bracket is necessary to ensure the K\"ahler invariance. For instance,
if we apply the K\"ahler transformation, $K \to K - \langle K_\phi
\rangle \phi - \langle K_{\phi^*} \rangle \phi^*$, the first term
vanishes and the second term compensates it. This becomes clear if we
write the interactions in terms of $G$. The effective Lagrangian
(\ref{eq:int-1}) is represented as
\beq
\mathcal{L} \;=\; - \frac{1}{2} e^{G/2} 
(G_{\phi ij} - 2 \Gamma^k_{\phi i} G_{jk}) 
\phi \chi^i \chi^j + {\rm h.c.},
\eeq
which is obviously invariant under the K\"ahler transformation.  Note
that $\Gamma^\ell_{\phi i}$ in Eq.~(\ref{eq:int-1}) is different from
$\Gamma^\phi_{ij}$ in Eq.~(\ref{hol-fermion}). The coefficient,
$\Gamma^k_{\phi i} \sim K_{\phi ik^*}$, can be nonzero easily. For
instance, $\delta K \sim |\phi|^2 |Q|^2$ leads to $\Gamma^Q_{\phi Q}
\sim \la \phi \ra$, which is nonzero as long as $\la \phi \ra \neq 0$.

On the other hand, the 2-scalar production consists of the two
channels; $\phi \to \phi^i \phi^j$ and $\phi \to \phi^i\phi^{*j}$. The
former comes from the scalar potential;
\begin{eqnarray}
    \mathcal{L} &=& 
    - \frac{1}{2} e^{K} 
    \left(
      K_\phi W_{ij} + W_{\phi ij}
      - 2 \Gamma^k_{\phi i} W_{jk}
    \right)^*
    g^{\phi\phi^*} W_{\phi\phi}\, 
    \phi \phi^{*i} \phi^{*j} + {\rm h.c.}.
    \label{eq:2scalar}    
\end{eqnarray}
We can easily check that this provides the same decay rate as that of
the fermion final state induced by (\ref{eq:int-1}), i.e.,
$\Gamma(\phi \to \bar\chi^i \bar\chi^j) = \Gamma(\phi \to \phi^i
\phi^j)$.  Also the kinetic term of the scalar fields gives another
decay channel, $\phi \to \phi^i\phi^{*j}$.  However its amplitude is
proportional to the scalar mass squared of the final state, noting
$\left[\phi |Q|^2\right]_D = \phi (\partial^2 Q^*) Q + \cdots$. Thus
the process becomes dominant only when the scalar fields has a quite
large soft scalar mass.

To summarize, the total decay rate of the 2-body final state from the
interactions (\ref{eq:int-1}) and (\ref{eq:2scalar}) is
\begin{eqnarray}
    \Gamma^{\rm (2-body)} &\equiv& 
    \Gamma(\phi \to \bar\chi^i \bar\chi^j) + 
    \Gamma(\phi \to \phi^i \phi^j) \non\\
    &\simeq& \frac{C^{(2)}_{ij}}{8\pi} m_{\phi} 
    \left( 1 - {4 M_Q^2 \over m_\phi^2} \right)^{\frac{1}{2}},
    \label{eq:2-body}
\end{eqnarray}
where $C^{(2)}_{ij} = e^K |K_\phi W_{ij} + W_{\phi ij} - 2
\Gamma^k_{\phi i} W_{jk}|^2$ with fixed $i$ and $j$ (the sum is taken
only over $k$). If the particles in the final state have a SUSY mass,
$W = M_Q QQ'$, $C^{(2)}_{ij}$ is proportional to $M_Q^2$. If the two
particles in the final state are identical to each other, $W =
\frac{1}{2} M_Q QQ$ (e.g. the right-handed neutrino $N$ with a
Majorana mass $M_N$), the decay rate becomes half of
(\ref{eq:2-body})~\footnote{The spontaneous decay into the
right-handed (s)neutrinos make the non-thermal leptogenesis scenario
quite attractive~\cite{Endo:2006nj}.}.

\begin{figure}[t!]
  \begin{center}
    \includegraphics[scale=1.2]{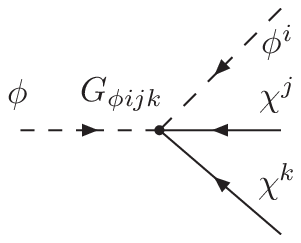}
    \includegraphics[scale=1.2]{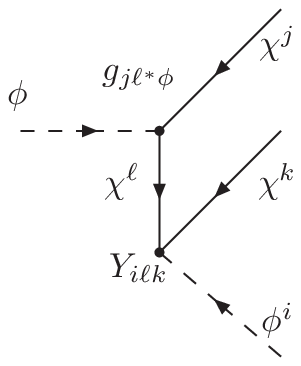}
    \includegraphics[scale=1.2]{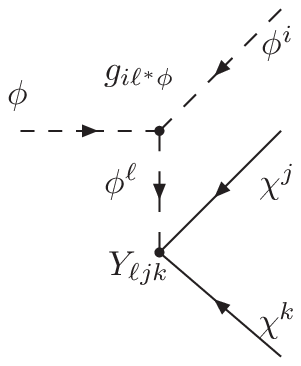}
    \caption{ The decay of the inflaton into the three-body 
      final states; the decay with the four-point vertex, and
      with the fermion and scalar exchanges, 
      from left to right. }
    \label{diagram1}
  \end{center}
\end{figure}

Next we consider the decay with 3-body final states.  The decay
processes through the dimension five operators are $\phi \to \phi^{*i}
\bar\chi^j \bar\chi^k$ and $\phi \to \phi^i \phi^j \phi^k$. The former
process is composed of the three diagrams in Fig.~\ref{diagram1}. In
addition to the spontaneous decay process pointed out in
\cite{Endo:2006qk} (the left diagram), the higher dimensional terms in
the K\"ahler potential contribute to the decay rate (middle and
right).  Evaluating these diagrams, we obtain the effective
interactions as
\begin{eqnarray}
    \mathcal{L} &=& -\frac{1}{2} e^{K/2} 
    \left(
      K_\phi W_{ijk} + W_{\phi ijk} 
      - 3 \Gamma^\ell_{\phi i} W_{jk\ell}
    \right)\, \phi \phi^i \chi^j \chi^k + {\rm h.c.}.
\end{eqnarray}
On the other hand, the interactions representing the decay into $3$
scalars, $\phi \to \phi^i \phi^j \phi^k$, are obtained by expanding
the scalar potential as
\begin{eqnarray}
    \mathcal{L} &=& -\frac{1}{6} e^{K} 
    \left(
      K_\phi W_{ijk} + W_{\phi ijk} 
      - 3 \Gamma^\ell_{\phi i} W_{jk\ell}
    \right)^* g^{\phi\phi^*} W_{\phi\phi}\, 
    \phi \phi^{*i} \phi^{*j} \phi^{*k} + {\rm h.c.}.
\end{eqnarray}
One can write down these interactions in terms of the K\"ahler
invariant function, $G$, by replacing $K_\phi W_{ijk} + W_{\phi ijk}
\to G_{\phi ijk}$ and $W_{jk\ell} \to G_{jk\ell}$, respectively. We
find that the decay rate into $3$ scalars is same as that into $1$
scalar + $2$ fermions, i.e., $\Gamma(\phi \to \phi^i \phi^j \phi^k)
\simeq \Gamma(\phi \to \phi^{*i} \bar\chi^j \bar\chi^k) + \Gamma(\phi
\to \phi^{*j} \bar\chi^k \bar\chi^i) + \Gamma(\phi \to \phi^{*k}
\bar\chi^i \bar\chi^j)$, for fixed $i$, $j$ and $k$. Summing these
decay rates, the total 3-body decay rate is given by
\begin{eqnarray}
    \Gamma^{\rm (3-body)} &\equiv& 
    \Gamma(\phi \to 3\,{\rm scalars}) + 
    \Gamma(\phi \to 1\,{\rm scalar} + 2\,{\rm fermions}) 
    \non\\
    &\simeq& 
    \frac{C^{(3)}_{ijk}}{256\pi^3} m_{\phi}^3,
    \label{eq:3-body}
\end{eqnarray}
where $C^{(3)}_{ijk} = e^K |K_\phi W_{ijk} + W_{\phi ijk} - 3
\Gamma^\ell_{\phi i} W_{jk\ell}|^2$ with fixed $i$, $j$ and $k$ (sum
over $\ell$). Here we have neglected the masses of the final-state
particles.

Finally we discuss the inflaton decay into the gauge bosons and
gauginos. At the tree level, it is effective only when the gauge
kinetic function depends on the inflaton field~\footnote{
  The coupling may be induced by the mixing of the inflaton with other
  fields such as the SUSY breaking field~\cite{Endo:2006tf}.
}.
Actually, we obtain the total rate of the decay into the gauge sector 
as~\cite{moduli}
\begin{eqnarray}
    \Gamma^{({\rm gauge~ tree})} &\simeq& 
    \frac{N_g}{4\pi} |\kappa|^2 m_\phi^3, 
    \label{eq:gauge-direct}
\end{eqnarray}
from $\mathcal{L} = \kappa \int d^2\theta\, \phi W^\alpha W_\alpha$,
where $W_\alpha$ is a field strength of the gauge supermultiplet,
$N_g$ is a number of the generators of the gauge symmetry, and we have
assumed the canonical normalization for the inflaton and gauge
multiplet. In (\ref{eq:gauge-direct}), half of the decay rate comes
from the gauge boson production and the other half is from the gaugino
production.

Except for such direct couplings, no sizable interactions are found at
the tree level between the inflaton and the gauge fields in the SUGRA
Lagrangian~\cite{WessBagger}. The feature can be understood by using
the gravity supermultiplet. The multiplet is minimally composed of the
followings;
\begin{eqnarray}
    h_{mn},~~\psi_m^\alpha,~~b_m,~~M,
\end{eqnarray}
which represent the graviton, the gravitino, and the vector and scalar
auxiliary fields that correspond to the $U(1)_R$ and conformal
symmetries of the superconformal transformation, respectively. Even in
the absence of the direct couplings, the gravity multiplet can connect
the inflaton field to the visible/hidden sectors. Actually, the
auxiliary fields, $b_m$ and $M$, depend on the inflaton field as well
as the visible/hidden fields due to the equation of motion, and the
longitudinal component of the graviton, $h$, is related to the
inflaton through the Lagrangian term, $\mathcal{L} = - \frac{1}{2}
e^{-K/3} \mathcal{R}$, in the conformal frame~\footnote{ Even in the
non-SUSY models, the latter contribution can
arise~\cite{Watanabe:2006ku}.  }.

The relevant terms involving $b_m$ in the SUGRA
Lagrangian~\cite{WessBagger} are given by
\begin{eqnarray}
    \mathcal{L}_{aux} = \frac{1}{3} b^m b_m
    - \frac{1}{3} i(K_i \partial_m \phi^i - 
    K_{i^*} \partial_m \phi^{*i}) b^m
    + \frac{1}{6} g_{ij^*} \bar\chi^j \bar\sigma_m \chi^i b^m
    - \frac{1}{2} \bar\lambda \bar\sigma_m \lambda b^m.
\end{eqnarray}
Solving the equation of motion for $b_m$, one can see that $b_m$
depends linearly on $\phi$ with a coefficient $K_\phi$, and that it
also includes the gaugino current, $\bar\lambda \bar\sigma_m
\lambda$. The decay into a pair of the gauginos is thus suppressed by
the gaugino mass because the processes requires a chirality flip. In
other words, noting that the inflaton contributes to the longitudinal
component of $b_m$, the $U(1)_R$ charges of the final state should be
nonzero for the decay to proceed due to the $U(1)_R$ current
conservation.  Thus the gaugino mass appears in the amplitude.

Next we focus on $h$ and $M$. The superconformal calculus formulation
of SUGRA~\cite{Cremmer:1978hn} is convenient to understand the decays
mediated by these fields. In fact, their contributions can be taken
into account by incorporating the chiral compensator field into the
Lagrangian. The F-term of the compensator corresponds to $M$ by using
the equation of motion, and $M$ includes a linear term with respect to
$\phi$, whose coefficient is proportional to $K_\phi$.  Further, since
the compensator has a Weyl charge, its scalar component depends on $K$
after the Weyl transformation to canonicalize the gravity sector,
i.e. from the conformal frame into the Einstein one. Then $\phi$
arises linearly in the scalar component when $K_\phi$ is
non-zero. Thus the operators induced by $h$ and $M$ are represented by
multiplying the compensator field. It is, however, known that the
compensator does not physically couple to the gauge sector because it
is conformal. Consequently, the decays into the gauge sector are
suppressed at the tree level.

Before closing, it is interesting to note that these features are
broken at the quantum level.  That is, the inflaton can decay into the
gauge sector via anomalies. We will discuss this mechanism in the next
section.

\section{Anomaly-induced Decay}
\label{sec:4}

At the classical level, the spontaneous decay of the inflaton into the
gauge sector is suppressed, since the gauge sector is conformal as
discussed in the previous section. The quantum corrections, however,
violate the conformal invariance, and so, the inflaton decay into the
gauge sector may arise at the quantum level. Taking account of the
SUGRA effects, the super-Weyl-K\"ahler (SW-K\"ahler) symmetry and the
sigma-model isometry are anomalous at the quantum level. Not only do
these anomalies mediate the SUSY-breaking effects to the visible
sector~\cite{AMSB}, but they also enable the inflaton field to couple
to the gauge supermultiplets~\cite{Endo:2007ih}.

In the superfield description, the 1PI effective action includes the
non-local terms corresponding to the
anomalies~\cite{LopesCardoso:1993sq,Bagger:1999rd};
\begin{eqnarray}
    \Delta\mathcal{L} &=& 
    - \frac{g^2}{(16\pi)^2} 
    \int d^2\theta\, W^\alpha W_\alpha 
    \frac{\bar{D}^2}{\partial^2}
    \bigg[ 
      4(T_R - 3T_G) R^\dagger
      \nonumber \\ && ~~~~~~~~~~~~~~~~~~~
      - \frac{T_R}{3} D^2 K
      + \frac{T_R}{d_R} D^2 \log\det K|_R^{''} 
    \bigg]
    + {\rm h.c.}
    \label{eq:non-local}
\end{eqnarray}
at the leading order of $1/M_P$ in the conformal frame. Here $D$ is a
covariant derivative of the supersymmetry, and $g$ is a gauge coupling
constant. The coefficients, $T_G$ and $T_R$, are the Dynkin index of
the adjoint representation and matter fields in the representation $R$
of dimension $d_R$, which are normalized to $N$ for SU($N$) and $1/2$
for its fundamentals. A sum over the matter fields is understood. Also
$K|_R^{''}$ denotes the K\"ahler metric restricted to the
representation $R$. The first term in the bracket of
Eq.~(\ref{eq:non-local}) corresponds to the SW anomaly, and it is not
invariant under the SW transformation.  In fact, the superspace
curvature $R$ changes under the SW transformation
as~\cite{WessBagger};
\begin{eqnarray}
    \delta R &=& -2 (2\Sigma - \bar\Sigma) R 
    - \frac{1}{4} \bar{D}^2 \bar\Sigma,
\end{eqnarray}
where a chiral superfield $\Sigma$ is defined so as to rescale the
vielbein, $\delta E_M^a = (\Sigma + \bar\Sigma) E_M^a$, and the last
term induces a shift of $R$. On the other hand, the second and third
terms in Eq.~(\ref{eq:non-local}) arise from the K\"ahler and
sigma-model anomalies, respectively.

In the conformal frame, $R^\dagger$ is expanded as $R^\dagger =
-\frac{1}{6} [M^* + \theta^2 (-{\cal R}/2 + i \partial_m b^m ) ] +
\cdots$~\cite{WessBagger}, where $\cdots$ is irrelevant for the
decay. In addition to the auxiliary fields, the Ricci scalar,
$\mathcal{R}$ depends on the inflaton field through the kinetic term,
$\mathcal{L} = - \frac{1}{2} e^{-K/3} \mathcal{R}$, which induces the
mixing of the inflaton with the longitudinal mode of the graviton.  To
simplify the calculation, let us go to the Einstein frame where the
gravity is canonically normalized. To this end, we perform the SW
transformation with $\Sigma_E = \phi_E + \sqrt{2} \theta \chi_E +
\theta^2 F_E$ defined by~\cite{Bagger:2000dh}
\begin{eqnarray}
    \phi_E = \frac{1}{12} K,~~
    \chi_E = \frac{1}{6} K_i\chi^i,~~
    F_E = \frac{1}{6} K_iF^i 
    - \frac{1}{12} K_{ij}\chi^i\chi^j.
\end{eqnarray}
Then the anomaly-induced term becomes~\cite{Bagger:2000dh}
\begin{eqnarray}
    \Delta\mathcal{L}_E = \Delta\mathcal{L} 
    + \frac{g^2}{16\pi^2} (T_R - 3T_G)
    \int d^2\theta\, \Sigma_E W^\alpha W_\alpha 
    + {\rm h.c.},
\end{eqnarray}
where the fields in $\Delta\mathcal{L}$ are simply replaced by those
defined in the Einstein frame~\footnote{ A factor in front of $T_R$ is
different from the result in \cite{Bagger:2000dh} because here $K$ in
$\Delta\mathcal{L}$ is not shifted.  }.

Expanding the superfields in terms of the components, one obtains
interaction terms of the inflaton field to the gauge
bosons/gauginos~\footnote{
  This result is also obtained at the component level by the Weyl
  rescaling, $e_m^a \to e^{-2\sigma} e_m^a$, from the conformal frame
  to the Einstein frame. Then the $\mathcal{R}$ and $M$ shift as
  $\delta \mathcal{R} = 12 \,\partial^2\, \sigma$ and $\delta M = -
  K_i F^i$ with $\sigma = K/12$, while $b_m$ remains unchanged.
};
\begin{eqnarray}
    \mathcal{L} &=& 
    \frac{g^2}{64\pi^2} X_G \;
    \phi ( F_{mn} F^{mn} -i F_{mn} \tilde F^{mn} )
    - \frac{g^2}{32\pi^2} X_G
    m_\phi \phi^* \lambda\lambda 
    + {\rm h.c.},
    \nonumber \\
    X_G &=& (T_G - T_R) K_\phi + \frac{2 T_R}{d_R} 
    (\log\det K|_R^{''})_{,\phi},
\label{eq:anom}
\end{eqnarray}
where $F_{mn}$ is a field strength of the gauge field and $\tilde
F^{mn} = \epsilon^{mnkl} F_{kl}/2$.  Here we have also used the
equations of motion for the auxiliary fields in the Einstein frame;
\begin{eqnarray}
    b_m = \frac{1}{2} i(K_i \partial_m \phi^i 
    - K_{i^*} \partial_m \phi^{*i}) + \cdots,~~~
    F^i = - e^{K/2} g^{ij^*} (W_j + K_j W)^*.
\end{eqnarray}
It is noticed that $M^* = -3 e^{K/2} W^*$ does not induce the decay
because of $|W_\phi| \sim m_{3/2} \la \phi \ra$ for the inflaton,
$\phi$.  The total decay rate from (\ref{eq:anom}) becomes
\begin{eqnarray}
    \Gamma^{({\rm anomaly})} &\simeq&
    \frac{N_g \alpha^2}{256 \pi^3} |X_G|^2 m_\phi^3, 
    \label{eq:anomaly-decay}
\end{eqnarray}
where $\alpha$ is a fine structure constant of the gauge group. 
Note that half of the decay rate comes from the decay into the two
gauge bosons, while the other half from that into the gaugino pair.

Let us compare the rate of the anomaly-induced decay
(\ref{eq:anomaly-decay}) with that of the spontaneous decay at the
tree level (\ref{eq:2-body}) and (\ref{eq:3-body}). We find that all
these rates are proportional to $|K_\phi|^2$. It means that, if the
K\"ahler potential of the inflaton is canonical, the VEV of the
inflaton field is necessary for the decay to proceed by the SUGRA
effects. In contrast, the phase space and coupling constants depends
on each process. The decay rate into the 2-body final state
(\ref{eq:2-body}) is suppressed by the mass squared, $M_Q^2/M_P^2 \ll
1$. While the rate of the 3-body final state (\ref{eq:3-body}) is
suppressed by the phase space compared to (\ref{eq:2-body}). Instead,
the coupling constant is given by the Yukawa coupling, $W_{ijk}$.
Compared to these tree-level processes, the anomaly-induced decay
takes place at the one-loop level. However, since the final state is 2
body, i.e. a pair of gauge bosons and gauginos, its rate is not
negligible compared to those of the spontaneous decays at the tree
level.

Let us comment on a mass spectrum of the matters in the visible/hidden
sectors. In this section, we have discussed anomalies that connects
the inflaton with the gauge sector. In order for the process to occur,
masses of the matters which contribute to the anomaly diagrams must be
smaller than the inflaton mass. Otherwise the matters decouple from
the anomalies. For instance, when we consider the anomaly-induced
decay into the SUSY breaking sector, since masses in the hidden quarks
are expected to be of $O(\Lambda)$, the decay takes place only for
$m_\phi > \Lambda$.

So far, we have considered the anomalies of the SW-K\"ahler symmetry
and sigma-model isometry. Since the process is an one-loop effect,
there may be possible contributions from the counter term, depending
on the underlying physics. Although we have assumed the conformal
frame without the counter term at the cutoff scale in the above
analyses, it can affect the decay rate, which is analogous to the
anomaly-mediated SUSY breaking scenario~\cite{Bagger:1999rd}.

Finally, let us comment on the inflaton decay into the SUSY breaking
sector which involves the conformal dynamics. If the inflaton mass is
above the scale of the violation of the conformal dynamics~\footnote{
See \cite{Ibe:2007wp} for a conformal theory of the SUSY breaking.  },
its decay into the SUSY breaking sector is expected to be
suppressed. Actually, since the beta function vanishes above the
scale, the decay induced by the SW anomaly is forbidden. At the same
time, the contributions from the K\"ahler and sigma-model anomalies
are implied to be suppressed~\footnote{ M.E. thanks K.-I.~Izawa for
discussions.  }, once we notice that the SUSY breaking sector is
sequestered from the other sectors by the conformal
dynamics~\cite{Luty:2001zv}.  Then the inflaton field may not decay
into the conformal SUSY breaking sector, and so, the models will be
free from the gravitino production.

\subsection{Minimal K\"ahler Potential}

Let us explicitly show several examples of the spontaneous and
anomaly-induced decays. The former decay depends on the K\"ahler
potential of the inflaton and visible/hidden sectors. Let us first
discuss the case of the minimal K\"ahler potential.  We take the
K\"ahler potential and the superpotential as
\begin{eqnarray}
    K &=& \phi\phi^* + QQ^*, \\
    W &=& W(\phi) + \frac{1}{2} M QQ 
    + \frac{1}{6} Y_{ijk} Q^i Q^j Q^k,
\label{eq:sp} 
\end{eqnarray}
where $Q$ denotes the visible/hidden matters~\footnote{ In addition,
  there may be a soft scalar mass in the K\"ahler potential.  However
  it is irrelevant for the spontaneous and anomaly-induced decay
  processes.  }.  Then the total decay rate is the sum of the
  spontaneous and anomaly-induced decays, $\Gamma = \Gamma^{({\rm
  tree})} + \Gamma^{({\rm anomaly})}$. The former is given by
\begin{eqnarray}
    \Gamma^{({\rm tree})} &\simeq& 
    \frac{N^{(2)}}{16\pi} 
    \frac{\langle \phi \rangle^2}{M_P^2}
    \frac{M^2}{M_P^2}
    m_\phi
    \left( 1 - \frac{4 M^2}{m_\phi^2} 
    \right)^{\frac{1}{2}}
    + 
    \frac{N^{(3)}}{256\pi^3} 
    \frac{\langle \phi \rangle^2}{M_P^2}
    |Y_{ijk}|^2
    \frac{m_\phi^3}{M_P^2}
    \label{eq:minimal-tree}
\end{eqnarray}
for fixed $i$, $j$ and $k$ with $i \neq j \neq k$.  Here $N^{(2)}$ and
$N^{(3)}$ denote a number of the final states. On the other hand, the
anomaly-induced decay depends on the gauge structure. The decay rate
is
\begin{eqnarray}
    \Gamma^{({\rm anomaly})} &\simeq&
    \frac{N_g \alpha^2}{256 \pi^3} (T_G - T_R)^2 
    \frac{\langle \phi \rangle^2}{M_P^2}
    \frac{m_\phi^3}{M_P^2}.
\label{eq:minimal-anomaly}
\end{eqnarray}
In the above results, we have assumed that the inflaton mass is
dominated by the SUSY-invariant mass term in the superpotential, and
we have neglected the masses of the final states for the 3-body decay
and the anomaly-induced decay.

\subsection{Sequestered K\"ahler Potential}
\label{sec:3b}

The next example is the K\"ahler potential with a sequestering form;
\begin{eqnarray}
    K &=& -3 \log \left[ 1 - \frac{1}{3}
      (\phi\phi^* + QQ^*) \right], 
\end{eqnarray}
with the superpotential (\ref{eq:sp}).  Noting $\langle\Gamma^k_{\phi
i}\rangle = (\langle\phi\rangle/3) \delta^k_i$, the rates of the
spontaneous and anomaly-induced decays are
\begin{eqnarray}
    \Gamma^{({\rm tree})}  &\simeq& 
    \frac{N^{(2)}}{144\pi} 
    \frac{\langle \phi \rangle^2}{M_P^2}
    \frac{M^2}{M_P^2}
    m_\phi
    \left( 1 - \frac{4 M^2}{m_\phi^2} 
    \right)^{\frac{1}{2}} \non\\
  \Gamma^{({\rm anomaly})} &\simeq&    
    \frac{N_g \alpha^2 b_0^2}{2304 \pi^3} 
    \frac{\langle \phi \rangle^2}{M_P^2}
    \frac{m_\phi^3}{M_P^2}.
    \label{eq:sequestered}
\end{eqnarray}
where $b_0$ is the beta function of the gauge symmetry, $b_0 = 3T_G -
T_R$. The tree-level decay arises via the mass term of $Q$, and the
anomaly-induced decay is due to the SW anomaly, while the spontaneous
decay via the Yukawa coupling vanishes. We also find that the
radiative effects associated to the K\"ahler and sigma-model anomalies
cancel with each other, which is analogous to the cancellation of the
AMSB effects to the gaugino mass~\cite{AMSB} .

The cancellation of the 3-body decay can be understood in the
conformal frame. As was explained above, a part of the spontaneous
decay is mediated by $b_m$ and the others are represented in terms of
the chiral compensator field, $\Phi$. The auxiliary field $b_m$
behaves as the gauge field of $U(1)_R$ of the superconformal symmetry.
At the tree-level, this symmetry is preserved by assigning an $U(1)_R$
charge $2/3$ for the chiral compensator. Then after a field
redefinition, $\Phi Q \to Q$, $U(1)_R$ charge vanishes for the
operators which represent the 3-body final state of the decay, that
is, $\phi^i\chi^j\chi^k$ and $W_{\phi\phi} \phi^i \phi^{*j}\phi^{*k}$.
Consequently, the decay mediated by $b_m$ cannot proceed via the
Yukawa interaction. The other tree-level processes induced by the
gravitational effects are also suppressed for the 3-body decay. They
are obtained by multiplying $\Phi$. Remembering that the
anomaly-mediated SUSY breaking contributions to the soft trilinear
couplings are absent at the tree level, $\Phi$ does not contribute to
the Yukawa interactions physically. Actually, $\Phi$ becomes absent in
the Yukawa interaction by the redefinition of the matte field, $\Phi Q
\to Q$.

In addition to the gravitational effects discussed above, the inflaton
decay may be induced by higher dimensional operators in the K\"ahler
function, $\Omega$, which is defined as $\Omega = -3 e^{-K/3}$. In
fact, in contrast to the sequestered K\"ahler potential, there exist
the higher dimensional terms in $\Omega$ for the minimal K\"ahler
potential. Thus the 3-body decays are allowed for the minimal case
(see (\ref{eq:minimal-tree})), while they are absent in the
sequestered one (see (\ref{eq:sequestered}))~\footnote{ In the
Einstein frame, the cancellation can be seen explicitly by the field
redefinition, $e^{\hat K/6} Q \to Q$, with $\hat K = K - \langle K
\rangle$. This rescaling substantially corresponds to the
transformation from the Einstein frame to the conformal one with
respect to the interaction terms of the matters.  }.

\section{Cosmology}
\label{sec:5}
We now consider cosmological implications of the inflaton decay
processes discussed in the preceding sections. One immediate
consequence is that the reheating temperature $T_R$ is bounded below;
$T_R$ cannot be arbitrarily low, since the inflaton decays into the
visible sector through the top Yukawa coupling (See
Eq.~(\ref{eq:3-body})).  The other is the gravitino production from
inflaton decay, which can occur through three different processes: (i)
gravitino pair production; (ii) spontaneous decay at the tree level;
(iii) anomaly-induced decay at the one-loop level. We will show how
severely the gravitino production constrains the inflation models and
SUSY breaking scenarios.

\subsection{Lower bound on the reheating temperature}
Let us begin with a relatively simple exercise.  The supersymmetric SM
sector contains the top Yukawa coupling in the superpotential as
\beq
W \;=\; Y_t \,T Q H_u,
\eeq
where $Y_t$ is the top Yukawa coupling, and $T$, $Q$, and $H_u$ are
the chiral supermultiplets of the right-handed top quark and
left-handed quark doublet of the third generation, and up-type Higgs,
respectively.  In this section, we assume that the inflaton has the
minimal K\"ahler potential for simplicity. The partial decay rate of
the inflaton through the top Yukawa coupling is then
\beq
\label{eq:rate-th-top}
\Gamma_T \;\simeq\; \frac{3}{128 \pi^3} |Y_t|^2 \la \phi \ra^2  m_\phi^3,
\eeq
where $\la \phi \ra$ and $m_\phi$ are VEV and the mass of the
inflaton, respectively. The partial decay rate (\ref{eq:rate-th-top})
is derived from Eq.~(\ref{eq:3-body}) by noting $C^{(3)} \simeq Y_t^2
|\la \phi \ra|^2$ and additional numerical factor $6$ coming from
SU(3) $\times$ SU(2).  The presence of the decay through the top
Yukawa coupling sets a lower bound on the reheating temperature,
$T_R$.  We define the reheating temperature as
\beq
\label{eq:def-Tr}
T_R \;\equiv\; \lrfp{\pi^2 g_*}{10}{-\frac{1}{4}} \sqrt{\Gamma_\phi},
\eeq
where $g_* $ counts the relativistic degrees of freedom, and
$\Gamma_\phi$ denotes the total decay rate of the inflaton.  Using
Eqs.~(\ref{eq:rate-th-top}) and (\ref{eq:def-Tr}), we obtain the lower
bound on $T_R$,
\beq
T_R \;\gtrsim\; 1.9 \times 10^3 {\rm\,GeV}\, |Y_t| 
\lrfp{g_*}{200}{-\frac{1}{4}} 
\lrf{\la \phi \ra}{10^{15}{\rm \,GeV}}
\lrfp{m_\phi}{10^{12}{\rm\,GeV}}{\frac{3}{2}}.
\label{eq:low-bound-on-TR}
\eeq

Similarly the inflaton decays into the gluons and gluinos via the
anomalies of SUGRA. One can estimate the decay rate from
Eq.~(\ref{eq:anomaly-decay}) as
\beq
\Gamma_{\rm SU(3)} \;\simeq\; \frac{9}{32 \pi^3} \alpha_s^2 \la \phi \ra^2  m_\phi^3,
\eeq
where $\alpha_s = g_s^2/4 \pi$ denotes the SU(3)$_C$ gauge coupling
constant. Substituting $\alpha_s \simeq 0.05$, we can see $\Gamma_{\rm
SU(3)}$ is one order of magnitude smaller than $\Gamma_T$.  Therefore
the spontaneous decay into the visible sector is dominated by that
through the top Yukawa coupling, unless the K\"ahler potential takes a
specific form, i.e. the sequestered type (see Sec.~\ref{sec:3b}).

\begin{figure}[t]
\begin{center}
\includegraphics[width=12cm]{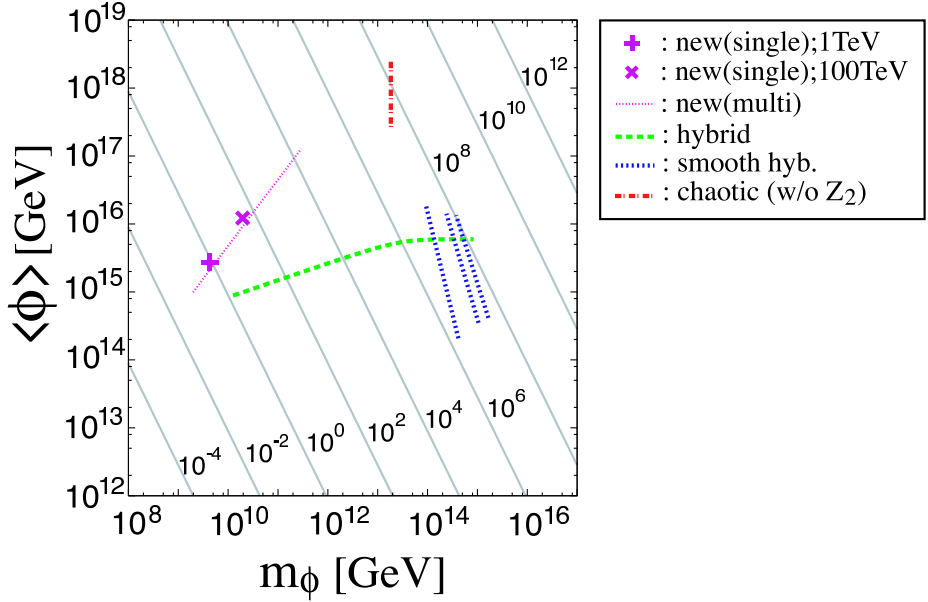}
\caption{Contours of the lower bound on $T_R$ in units of GeV. 
We set $g_* \;=\; 228.75$ and $Y_t \;=\; 0.6$.  
For details of the models, see Sec.~\ref{sec:models}. }
\label{fig:contour}
\end{center}
\end{figure}

We show the contours of the lower limit on $T_R$ given by
Eq.~(\ref{eq:low-bound-on-TR}) in Fig.~\ref{fig:contour}, together
with typical values of $\la \phi \ra$ and $m_\phi$ for the
single-field new~\cite{Izawa:1996dv}, multi-field
new~\cite{Asaka:1999jb}, hybrid~\cite{Copeland:1994vg} and smooth
hybrid~\cite{Lazarides:1995vr}, and chaotic~\cite{Kawasaki:2000yn}
inflation models. We will discuss each inflation model later in this
section.  If the inflaton mass $m_\phi$ and the VEV $\la \phi \ra$ are
too large, the reheating temperature may exceed the upper bound from
the gravitinos produced by particle scattering in the thermal
plasma. The cosmological constraints on the gravitino are summarized
in Sec.~\ref{sec:constraints-on-gravitino}. For more details, the
reader should refer to Refs.~\cite{Kawasaki:2004yh, Kohri:2005wn,
Jedamzik:2006xz} for the unstable gravitino, and
Refs.~\cite{Moroi:1993mb,Bolz:1998ek,Bolz:2000fu,Ellis:2003dn,Steffen:2006hw,Pradler:2006qh}
for the stable one.  For instance, the reheating temperature is
necessarily higher than $10^6{\rm\,GeV}$ for the smooth hybrid
inflation model, which is difficult to be reconciled with the
gravitino of $m_{3/2} \;=\; O(0.1-1){\rm\,TeV}$~\cite{Kawasaki:2004yh}
and $ 10{\rm\,eV} \lesssim m_{3/2} \lesssim
10{\rm\,MeV}$~\cite{Moroi:1993mb}.

It is remarkable that the inflaton decays into the visible sector once
it acquires a finite VEV; we do not need to introduce any interactions
between the inflaton and the SM sector by hand in the Einstein
frame~\footnote{ Note that the interpretation of higher dimensional
operators depends on a choice of the frame of SUGRA.  } in order to
induce the reheating. On the other hand, it may pose a cosmological
problem at the same time.  If the hidden sector also has a Yukawa
coupling or includes the SW-K\"ahler/sigma-model anomalies, unwanted
relics such as the gravitino may be directly produced by the inflaton.
We will focus on the issue in the rest of this section.

\subsection{Gravitino Production}

We consider the gravitino production from the inflaton decay. To make
our analysis simple and conservative, we assume that the inflaton has
the minimal K\"ahler potential and does not have any direct couplings
with the SUSY breaking sector in the superpotential.  If we introduce
possible couplings between the inflaton and the SUSY breaking field,
the gravitino overproduction problem generically becomes severer.  We
also assume the DSB scenario with the dynamical scale $\Lambda$.  Then
the SUSY breaking field $z$ usually has a scalar mass $m_z$ that is
much larger than the gravitino mass. Although the precise value of
$m_z$ is model-dependent, it is expected to be of the order of
$\Lambda$.  Hereafter we simply assume~\footnote{ The scalar mass
$m_z$ can be smaller than $\Lambda$.  If this is the case, the
pair-gravitino production will be affected.  }
\beq
m_z \simeq \Lambda \simeq \sqrt{m_{3/2}}.
\label{eq:z-mass}
\eeq
We discuss the cases of $m_\phi > \Lambda$ and $m_\phi < \Lambda$ 
separately.

\subsubsection{The case of $m_\phi < \Lambda$}
As we have seen in Sec.~\ref{sec:2}, the inflaton decays into a pair of the
gravitinos.  The gravitino pair production is effective especially for 
a low-scale inflation model with $m_\phi < m_z$.  The gravitino production 
rate is given by
\beq
\Gamma_{3/2}^{(\rm{pair})} \;\simeq\; \frac{1}{32 \pi} \la \phi \ra^2 m_\phi^3
\eeq
for $m_\phi < m_z \simeq \Lambda$. Here we have assumed the minimal K\"ahler 
potential with a soft scalar mass of $z$ and $\langle z \rangle \ll 1$.  
The gravitino abundance is then
\bea
Y_{3/2} &=& 2 
\frac{\Gamma_{3/2}^{(\rm{pair})}}{\Gamma_\phi} 
\frac{3 T_R}{4 m_\phi},\non\\
&\simeq&7 \times 10^{-11} 
\lrfp{g_*}{200}{-\frac{1}{2}} 
\lrfp{T_R}{10^6{\rm GeV}}{-1}
\lrfp{\la\phi\ra}{10^{15}{\rm GeV}}{2}
\lrfp{m_\phi}{10^{12}{\rm GeV}}{2}.
\label{Y32-pair} 
\eea
It should be noted that the gravitino abundance is inversely
proportional to $T_R$. This feature is to be contrasted to the
thermally produced gravitinos, whose abundance is proportional to
$T_R$.

\subsubsection{The case of $m_\phi > \Lambda$}
When the inflaton mass $m_\phi$ is larger than $\Lambda$, the
gravitational effects discussed in Sec.~\ref{sec:3} and \ref{sec:4}
are important. If the SUSY breaking sector has Yukawa interactions,
the inflaton decays into the sector via the operators. Besides, the
anomalies of SUGRA induce the inflaton decay into the gauge boson and
gauginos of the hidden gauge symmetries. Thus the hidden quarks and
gauge bosons/gauginos are generally produced at the decay for $m_\phi
> \Lambda$.

The hidden particles are energetic at the moment when they are
produced. Since the reheating temperature $T_R$ is bounded as $T_R <
\Lambda$ for almost entire region of the gravitino mass due to the
thermal-gravitino production, the produced hidden particles do not
reach thermal equilibrium. They instead form jets and hadronize by the
strong gauge interactions, followed by cascade decays of the heavy
hidden hadrons into lighter ones. The number of the hidden hadrons
produced from each jet, which we call here as the multiplicity $N_H$,
depends on the detailed structure of the hidden sector such as the
gauge groups, the number of the matter multiplets, and a mass spectrum
of the hidden hadrons. We expect $N_H$ to be in the range of $O(1 -
10^2)$.

The hidden hadrons should eventually decay and release their energy
into the visible sector, since otherwise they will easily overclose
the universe. The gravitinos are likely to be produced in the decays
of the hidden hadrons as well as in the cascade decay processes in
jets.  This happens, e.g. through the kinetic mixings of the hidden
matters, and especially if $z$ is a bound state of the hidden
(s)quarks.  Note that the goldstino is massless in the global SUSY
limit and it is in the hidden sector with renormalizable couplings to
other hidden (s)quark/gauge fields (and therefore hadrons). Thus, the
goldstinos are expected to be produced by the hidden hadrons, though
the precise production rate depends on details of the hidden
sector. We denote the averaged number of the gravitinos produced per
each jet as $N_{3/2}$. Here we assume each hidden hadron produces one
gravitino in the end, and use the relation $N_{3/2} \sim
N_H$~\footnote{ In particular, if $z$ is an elementary field and has a
Yukawa coupling, the inflaton necessarily produces at least one
goldstino by the decay through the coupling.  }.

The partial decay rates of the inflaton into the SUSY breaking sector
are given by Eqs.~(\ref{eq:minimal-tree}) and
(\ref{eq:minimal-anomaly}).  Although the DSB models do not always
possess Yukawa
interactions~\cite{Affleck:1983vc,Affleck:1984uz,Murayama:1995ng}, all
the DSB scenarios necessarily involve the gauge interactions.  From
Eq.~(\ref{eq:minimal-anomaly}), the partial rate of the inflaton decay
into the SUSY breaking sector is:
\beq
\Gamma_{\rm DSB} \;=\; \frac{N_g^{(h)} \alpha_h^2}{256 \pi^3} 
(T_G^{(h)}  - T_R^{(h)} )^2 \la \phi\ra^2 m_\phi^3,
\eeq
where the gauge coupling, the Dynkin indices, and the number of the
generators are those of the hidden gauge symmetries. Multiplying the
number of jets and $N_{3/2}$, the gravitino abundance becomes
\bea
Y_{3/2} & = &
 2 N_{3/2} \frac{\Gamma_{\rm DSB}}{\Gamma_\phi} \frac{3 T_R}{4 m_\phi},\non\\
 &\simeq& 9 \times 10^{-13} \xi \lrfp{g_*}{200}{-\frac{1}{2}} 
 \lrfp{T_R}{10^6{\rm GeV}}{-1}
 \lrfp{\la\phi\ra}{10^{15}{\rm GeV}}{2}
 \lrfp{m_\phi}{10^{12}{\rm GeV}}{2},
\label{Y32-anomaly} 
\eea
where we have defined $\xi \equiv N_{3/2} N_g^{(h)} \alpha_h^2 (T_G-T_R)^2$,
which is roughly expected to be in the range of $O(10^{-2})$ to $O(10)$.

In the following numerical analysis, we take the anomaly-induced decay
as a source of the gravitino production channel for $m_\phi >
\Lambda$.  As one can see from Eqs.~(\ref{eq:minimal-tree}) and
(\ref{eq:minimal-anomaly}), the decay rate is roughly comparable to
that of the spontaneous decay via the Yukawa coupling. Thus if one
includes the tree-level decay into the analysis, the constraints
become severer slightly, and the results in the followings do not
change essentially.

\subsection{Cosmological Constraints on Gravitinos}
\label{sec:constraints-on-gravitino}

Before going further, here we briefly summarize the cosmological constraints 
on the gravitinos, which will be used to put constraints on the inflation models 
later.

There are tight constraints on the gravitino abundance from BBN if the
gravitino is unstable~\cite{Kawasaki:2004yh, Kohri:2005wn,
Jedamzik:2006xz}~\footnote{ For early works, see
Refs.~\cite{BBNwX_OLD,Kawasaki:1994af, Protheroe:dt,
Holtmann:1998gd,Jedamzik:1999di,Kawasaki:2000qr,
Kohri:2001jx,Cyburt:2002uv}.}, and from the dark matter (DM) abundance
for the stable
gravitino~\cite{Moroi:1993mb,Bolz:1998ek,Bolz:2000fu,Ellis:2003dn,Steffen:2006hw,Pradler:2006qh}.
The abundance of the gravitinos produced by thermal scatterings is
related to $T_R$ as~\cite{Bolz:2000fu,Kawasaki:2004yh}
\begin{eqnarray}
    \label{eq:Yx-new}
    Y_{3/2}^{(th)} &\simeq& 
    1.9 \times 10^{-12}\left[ 1+ 
    \left(\frac{m_{\tilde{g}_3}^2}{3m_{3/2}^2}\right)\right]
    \left( \frac{T_{\rm R}}{10^{10}\ {\rm GeV}} \right)
    \nonumber \\ 
    & \times & 
    \left[ 1 
        + 0.045 \ln \left( \frac{T_{\rm R}}{10^{10}\ {\rm GeV}} 
        \right) \right]
    \left[ 1 
        - 0.028 \ln \left( \frac{T_{\rm R}}{10^{10}\ {\rm GeV}} ,
        \right) \right],
\end{eqnarray}
where we have taken $N=3$ for QCD and $m_{\tilde{g}_3}$ is the gluino
running mass evaluated at $T=T_R$. Since the gravitino abundance
$Y_{3/2}^{(th)}$ is roughly proportional to $T_R$, $T_R$ is bounded
from above.

Here we simply quote the bounds on $Y_{3/2}$ and $T_R$ summarized in
Ref.~\cite{Kawasaki:2006gs}.  If the gravitino is light, it is likely
the lightest SUSY particle (LSP) and therefore stable with the
R-parity conservation.  The bounds on $Y_{3/2}$ (and $T_R$) then come
from the requirement that the gravitino abundance should not exceed
the present DM abundance~\footnote{The gravitinos non-thermally
produced by the inflaton decay can be a dominant component of DM, for
certain values of the inflaton parameters~\cite{Takahashi:2007tz}.}:
\beq
m_{3/2} \,Y_{3/2} \;\leq\; \frac{\rho_c}{s} \Omega_{\rm DM}  \;\lesssim\;
4.4 \times 10^{-10} {\rm\,GeV},
\label{eq:gra-abu}
\eeq
where $\rho_c$ is the critical density, and we used $\Omega_{\rm DM}
h^2 \lesssim 0.12$ at $95\%$ C.L.~\cite{Spergel:2006hy} in the second
inequality. The upper bound on $T_R$ can be obtained by substituting
\EQ{eq:gra-abu} into \EQ{eq:Yx-new} as
\bea
\label{eq:stable-g2}
  T_R \; \lesssim \; \left\{
  \bear{ll}
  O(100)~{\rm GeV} &
  {\rm for}~~~m_{3/2}\simeq 10^{-2} - 10^{2}~{\rm keV}\\
  \ds{8 \times 10^7~{\rm GeV} 
    \lrfp{m_{\tilde{g}_3}}{300{\rm\,GeV}}{-2}
    \left(\frac{m_{3/2}}{1{\rm \,GeV}}\right)}
  &{\rm for}~~~m_{3/2}\simeq    10^{-4} - 10^2~{\rm GeV}
\eear  \right. .
\eea
Note that we have conservatively neglected the contribution from the
decay of the next-to-lightest SUSY particle.  In the following
analysis, we assume that the gravitino with a mass lighter than
$10^2$GeV is the LSP and stable.  When the gravitino is as light as
$m_{3/2} \sim O(10)$~eV~\cite{Viel:2005qj}, there are no constraints
on $T_R$, since the energy density of the gravitino would be too small
even if the gravitino is thermalized.

On the other hand, if the gravitino is unstable, BBN puts severe
constraints on $Y_{3/2}$~\cite{Kawasaki:2004yh,Kohri:2005wn}:
\begin{eqnarray}
\label{eq:unstable-Y1}
   Y_{3/2}  & ~ \lesssim & \left\{\begin{array}{lcl}
   ~1\times 10^{-16} - 6\times 10^{-16}
   &{\rm for}    &  m_{3/2} \simeq 0.1 - 0.2~{\rm TeV} \\
   ~4\times 10^{-17} - 6\times 10^{-16}
   &{\rm for}    &  m_{3/2} \simeq 0.2 - 2~{\rm TeV} \\
   ~ 7 \times 10^{-17} - 2\times 10^{-14} 
   &{\rm for}    & m_{3/2} \simeq 2 - 10~{\rm TeV} \\
   ~ 6\times 10^{-13} - 2\times 10^{-12} 
   &{\rm for}    & m_{3/2} \simeq 10 - 30~{\rm TeV} \end{array}\right.
   ~~(B_h \simeq 1),\\[1em]
\label{eq:unstable-Y2}
   Y_{3/2}  &~ \lesssim & \left\{\begin{array}{lcl}
   ~1\times 10^{-16} - 5\times 10^{-14}
   &{\rm for}&  m_{3/2} \simeq 0.1 - 1~{\rm TeV} \\
   ~ 2\times 10^{-14} - 5\times 10^{-14}
   &{\rm for}&  m_{3/2} \simeq 1 - 3~{\rm TeV} \\ 
    ~ 3\times 10^{-14} - 2\times 10^{-13}
   &{\rm for}&  m_{3/2} \simeq 3 - 10~{\rm TeV} \end{array}\right.
   ~~(B_h \simeq 10^{-3}). 
\end{eqnarray}
The corresponding upper bounds on $T_R$ are
\bea
\label{eq:rtemp_0}
   T_R\; \lesssim\; \left\{ 
   \bear{ll} (1-4)\times 10^6~{\rm GeV} &{\rm
   for} ~~ m_{3/2} \simeq 0.1 - 0.2~{\rm TeV} \\ 3 \times 10^{5} - 4
   \times 10^{6}~{\rm GeV} &{\rm for} ~~ m_{3/2} \simeq 0.2 - 2~{\rm
   TeV} \\ 5\times 10^{5} - 1\times 10^{8}~{\rm GeV} & {\rm for} ~~
   m_{3/2} \simeq 2 - 10~{\rm TeV} \\ (3 -10)\times 10^9~{\rm GeV}
   &{\rm for} ~~ m_{3/2} \simeq 10 - 30~{\rm TeV} 
\eear \right.   ~~(B_h \simeq 1),\\[1em]
\label{eq:rtemp_1}
   T_R\; \lesssim\; \left\{ 
   \bear{ll} 1\times 10^{6}- 3\times 10^{8}~{\rm GeV} &{\rm
   for} ~~ m_{3/2} \simeq 0.1 - 1~{\rm TeV} \\ 
  (1-3)\times 10^8~{\rm GeV} &{\rm for} ~~ m_{3/2} \simeq 1- 3~{\rm TeV} \\ 
 2\times 10^{8} -  1\times  10^{9}~{\rm GeV} & {\rm for} ~~
   m_{3/2} \simeq 3 - 10~{\rm TeV}
 \eear \right.(B_h \simeq 10^{-3}). 
\eea
For the heavy gravitino of mass $\gtrsim 30 (10)$ TeV, no stringent
constraints are obtained from BBN.  However, another constraint comes
from the abundance of the LSP produced by the gravitino decay. Since
the gravitino life time is rather long, the produced LSPs will not
annihilate with each other.  Thus the upper bounds on $Y_{3/2}$ and
$T_R$ read
\beq
\label{Yh}
m_{\rm LSP} \,Y_{3/2}  \;\lesssim\;
4.4 \times 10^{-10} {\rm\,GeV},
\eeq
and
\beq
T_R \;\lesssim\; 2.5 \times 10^{10} 
\lrfp{m_{\rm LSP}}{100{\rm\,GeV}}{-1} {\rm GeV},
\label{eq:const-from-lsp}
\eeq
where $m_{\rm LSP}$ denotes the mass of the LSP.

As is well-known, all the above constraints have been usually applied
for the gravitinos from the thermal production. Since the gravitinos
are also non-thermally produced in inflaton decay, we obtain further
constraints on $T_R$, $m_{3/2}$, $\la \phi \ra$ and $m_\phi$ by
requiring the abundance of the non-thermally produced gravitinos
(\ref{Y32-pair}) and (\ref{Y32-anomaly}) to satisfy
(\ref{eq:gra-abu}), (\ref{eq:unstable-Y1}), (\ref{eq:unstable-Y2}), or
(\ref{Yh}). As we will see later, these new constraints drive (some
part of) the high-scale inflation models and the gravity mediation
into a corner.

\subsection{Constraints on Inflation Models and SUSY breaking}
\label{sec:constraints-on-models}
 Now we would like to derive constraints on the inflation and SUSY
breaking models, using the non-thermal production of the gravitinos
discussed above together with the thermal process. 

\begin{figure}[t]
\begin{center}
\includegraphics[width=12cm]{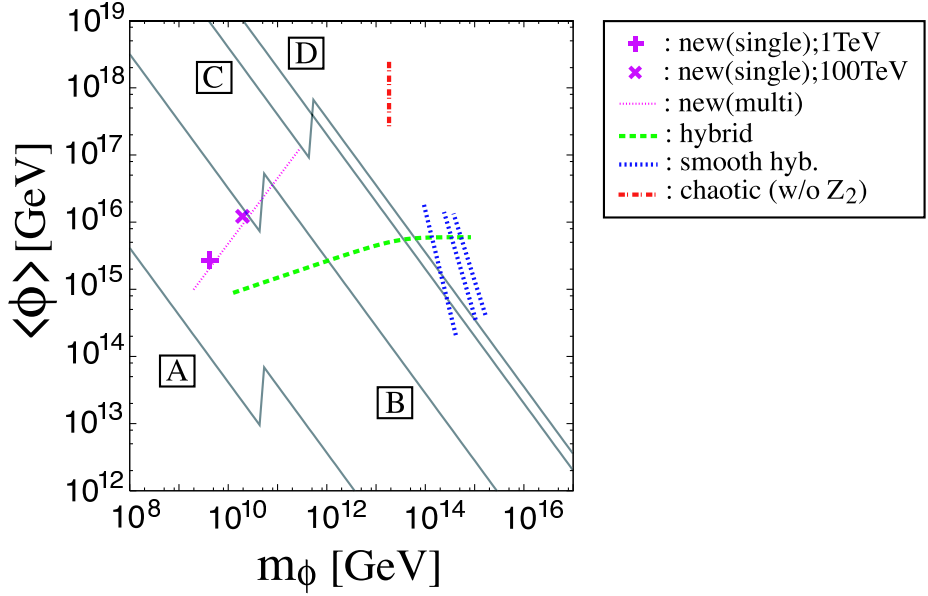}
\caption{ Constraints from the gravitino production by the inflaton
decay, for $m_{3/2} = 1{\rm\,TeV}$ with $B_h = 1$ (case\,A), $m_{3/2}
= 1{\rm\,TeV}$ with $B_h = 10^{-3}$ (case\,B), $m_{3/2} =
100{\rm\,TeV}$ (case\,C), and $m_{3/2} = 1{\rm\,GeV}$ (case\,D). The
region above the solid (gray) line is excluded for each case. 
For $m_\phi \gtrsim \Lambda$, we have used the anomaly-induced inflaton 
decay into the hidden gauge/gauginos to estimate the gravitino abundance, 
while the gravitino pair production has been used for $m_\phi \lesssim 
\Lambda$. Since $T_R$ is set to be the highest allowed value, the 
constraints shown in this figure are the most conservative ones.
}
\label{fig:m-vev}
\end{center}
\end{figure}

In Fig.~\ref{fig:m-vev}, we show the constraints on the inflaton mass
and VEV for $m_{3/2} = 1{\rm\,GeV}, 1{\rm\,TeV},$ and $100\,$TeV,
together with typical values of the inflation models. We discuss each
model in the next subsection. The region above each solid line is
excluded. We find that in the case of $m_{3/2} = 1$\,TeV with $B_h
=1$, all the inflation models shown in the figure are excluded. For
the gravitino mass lighter or heavier than the weak scale, the
constraints become relaxed. The inflaton mass and its VEV depend on
the inflation models.  Generically speaking, for larger $m_\phi$ and
$\la \phi \ra$, the constraints become severer, simply because more
gravitinos are produced by the inflaton decay (see (\ref{Y32-pair})
and (\ref{Y32-anomaly})).  On the other hand, if the inflaton is
charged under some symmetries, its VEV becomes suppressed or even
forbidden especially when the symmetry is exact at the vacuum. Then
the bounds can be avoided for such inflation models. This is the case
of the chaotic inflation model with a discrete symmetry (note that the
chaotic inflation model shown in Fig.~\ref{fig:m-vev} is that without
such a symmetry).

The solid lines which denote the constraint are jaggy at an
intermediate value of $m_\phi$. This is because the dominant
production channels of the gravitinos changes. In the right side, the
gravitinos are produced by the spontaneous and anomaly-induced decays,
while the inflaton directly decays into a pair of the gravitinos in
the left side. Note that we have assumed (\ref{eq:z-mass}) and $\xi =
1$ for simplicity.

In Fig.~\ref{fig:m-vev}, we have set $T_R$ to be the highest value
allowed by the constraints. As mentioned before, the abundance of the
non-thermally produced gravitinos is inversely proportional to $T_R$,
which is different from that of the thermally produced one (see
(\ref{eq:Yx-new})). If $T_R$ takes a smaller value, the constraints
becomes severer.  Thus, the bounds shown in Fig.~\ref{fig:m-vev} are
the most conservative ones. Note that one may have to introduce
couplings of the inflaton with the SM particles to realize the highest
allowed reheating temperature.

\begin{figure}[t]
\begin{center}
\includegraphics[width=14cm]{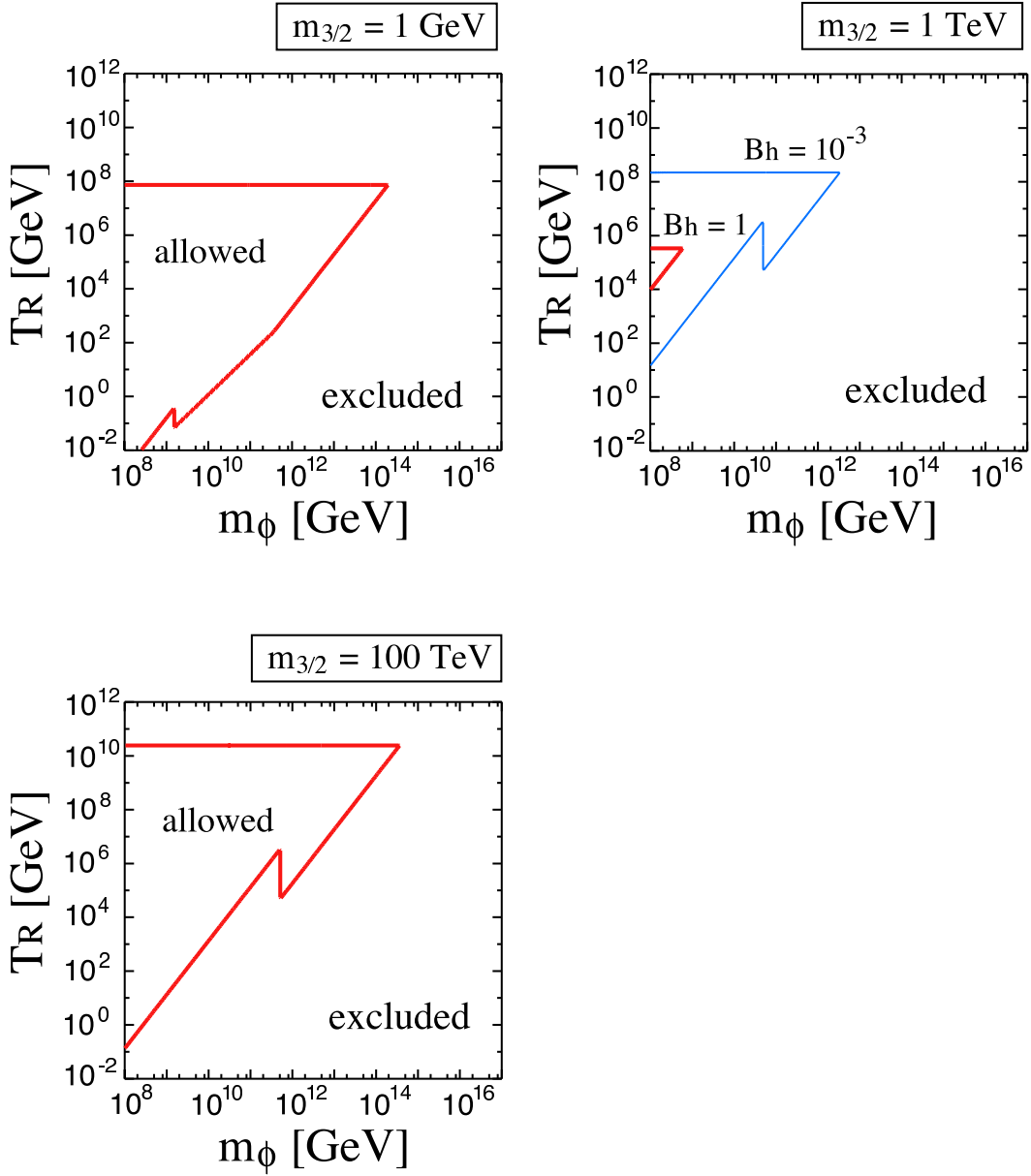}
\caption{ Constraints from the gravitino production by the inflaton
decay, for $m_{3/2} = 1{\rm\,GeV}$ (left-upper), $m_{3/2} =
1{\rm\,TeV}$ with $B_h = 1$ and $10^{-3}$ (right-upper), $m_{3/2} =
100{\rm\,TeV}$ (bottom). We have set $\la \phi \ra = 10^{15}$GeV.  The
region surrounded by the solid line is allowed for each case.  }
\label{fig:m-TR}
\end{center}
\end{figure}

Instead, taking $T_R$ as a free parameter, we show the constraints on
the $m_\phi - T_R$ plane, for $m_{3/2} = 1{\rm\,GeV}, 1{\rm\,TeV},$
and $100\,$TeV with a fixed $\la \phi \ra = 10^{15}{\rm\,GeV}$ in
Fig.~\ref{fig:m-TR}. The reheating temperature is bounded from above
due to the thermal production of the gravitino. It is remarkable that
we have lower bounds on $T_R$ due to the non-thermal processes.  In
the figure, we incorporated the spontaneous decay via the top Yukawa
interaction, which also provides a lower bound on $T_R$ (see
(\ref{eq:low-bound-on-TR})). One can see that the lower bound on $T_R$
becomes severer for larger $m_\phi$.

\begin{figure}[t]
\begin{center}
\includegraphics[width=14cm]{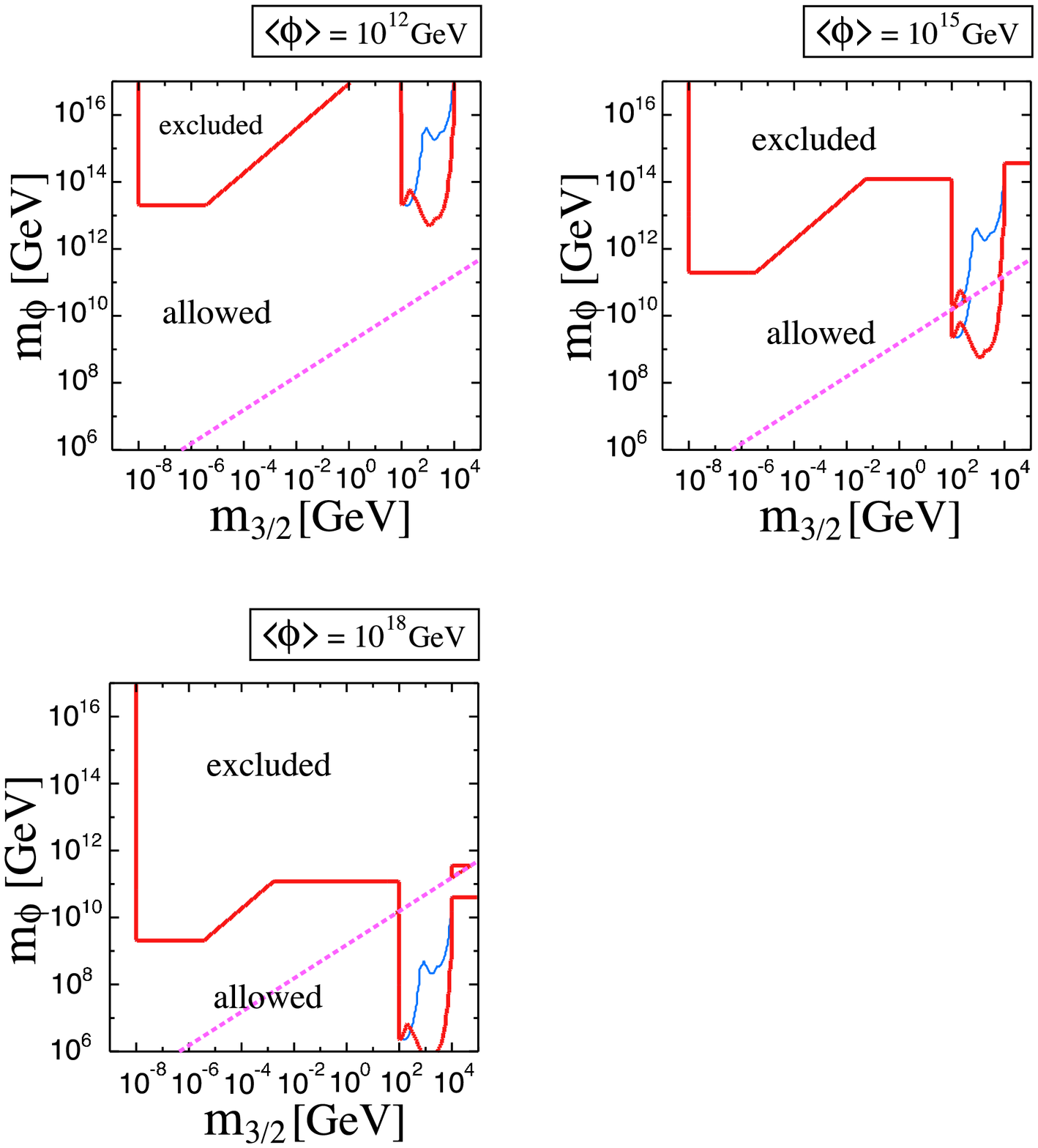}
\caption{ Constraints from the gravitino production by the inflaton
decay, for $\la \phi \ra = 10^{12}, 10^{15}$ and $10^{18}$GeV.  The
region above the thick solid line is excluded.  We also show the
constraint for the unstable gravitino with $B_h = 10^{-3}$ as the thin
(blue) line.  For the region above the dashed (pink) line, we adopt
(\ref{Y32-anomaly}), while (\ref{Y32-pair}) is used for the region
below the dashed line.  Since $T_R$ is set to be the highest allowed
value, the constraints shown in this figure are the most conservative
ones.  }
\label{fig:mg-m}
\end{center}
\end{figure}

In Fig.~\ref{fig:mg-m}, we show constraints on the $m_{3/2} - m_\phi$
plane for several values of $\la \phi \ra = 10^{12},\,10^{15},$ and
$10^{18}{\rm\,GeV}$.  The dashed (pink) line represents $m_\phi =
\Lambda$.  For the inflaton mass $m_\phi$ above the dashed (pink)
line, the spontaneous and anomaly-induced decays of the inflaton
produce the gravitinos, while the pair production is dominant below
the dashed (pink) line. We have set $T_R$ to be the highest value
allowed by the constraints as we did in Fig.~\ref{fig:m-vev}. We find
that the inflaton mass cannot be too large, especially for $m_{3/2}$
around the weak scale.  It is also noticed that the constraint becomes
severer as $\la \phi \ra$ increases, since the upper bound on $m_\phi$
is proportional to $\la \phi \ra^{-1}$ for fixed $m_{3/2}$.

Finally let us illustrate how much the problem becomes severer in the 
case of the SUSY breaking models with an elementary singlet $z$. 
Such a field is needed to give sizable masses to the gauginos in the 
simple version of the gravity-mediation~\cite{ModuliProblem,Banks:1993en} 
(see also footnote \ref{footnote:gravity-med}). 
For an inflaton with $m_\phi < \Lambda$, the pair gravitino production 
occurs as described before. In particular, since $z$ is singlet 
at the cutoff scale, there is a priori no reason to forbid such 
an interaction as $\delta K \sim |\phi|^2 (z+z^*)$. Then there 
generically exists a large kinetic mixing with the inflaton, and 
so, the gravitino production rate becomes too large, which is 
given by Eq.~(\ref{eq:pair-grav2}) with $\tilde{c} \sim 1$. With 
such a large gravitino production rate, most of the inflation 
models with $m_\phi < \Lambda$ are excluded, e.g. unless the 
inflaton VEV vanishes due to some symmetries. Even for $m_\phi > 
\Lambda$, the pair gravitino production occurs effectively. 
In fact, one expects that $\delta K \sim |\phi|^2 z z /2 + {\rm h.c.}$
generally exists. The gravitino pair production is then given by
(\ref{eq:pair-grav}) with $c \sim 1$. Thus the gravitino abundance
increases by $O(10^2)$ compared to that from the spontaneous and
anomaly-induced decays with $\xi = 1$ for $m_\phi > \Lambda$ (see
(\ref{Y32-pair}) and (\ref{Y32-anomaly}))~\footnote{ The spontaneous
decay at the tree level and the anomaly-induced one into the SUSY
breaking sector are not much affected by the presence of such an
elementary singlet $z$.  }.  Besides, such a $z$ field may be
displaced away from its potential minimum during the inflation,
forcing the cosmological scenario to be more problematic (see the
footnote~\ref{footnote:Polonyi}).  Thus the SUSY breaking models with
the elementary singlet $z$ are strongly disfavored from the
cosmological points of view.

\subsection{Inflation Models}
\label{sec:models}
 In this subsection we give a brief review on the representative
inflation models plotted in Figs.~\ref{fig:contour} and \ref{fig:m-vev}.  
For details on the models, the readers should refer to the original literatures.

\subsubsection{Single-field inflation model}
\label{sec:single}
As a concrete example, here we study the new inflation
model~\cite{Kumekawa:1994gx,Izawa:1996dv,Ibe:2006fs}.  In the new
inflation model, the K\"ahler potential and superpotential of the
inflaton sector are written as~\footnote{
The gravitino abundance in the text remains virtually unchanged in
the presence of the quartic coupling in the K\"ahler potential.
}
\bea
K(\phi,\phi^\dag) &=& |\phi|^2 + \frac{k}{4} |\phi|^4,\non\\
W(\phi) &=&  v^2\phi -  \frac{g}{n+1}\, \phi^{n+1}.
\eea
where the observed density fluctuations are explained for $v = 4\times
10^{-7} \, (0.1/g)^{1/2}$ and $k \lesssim 0.03$ in the case of
$n=4$~\cite{Ibe:2006fs}.  After inflation, the inflaton $\phi$ takes
the expectation value $\la\phi\ra\simeq (v^2/g)^{1/n}$. In this model
the inflaton mass is given by $m_{\phi} \simeq n v^2/\la\phi\ra$, and
the gravitino mass is related to $v$ as $m_{3/2} \simeq n v^2
\la\phi\ra /(n+1)$, since the inflaton induces the spontaneous
breaking of the $R$-symmetry.

In the case of $n=4$, the inflaton parameters are $m_\phi \simeq 4
\times 10^9$ GeV and $\la \phi \ra \simeq 3 \times 10^{15}$ GeV for
$m_{3/2}=1$~TeV, while $m_\phi \simeq 2 \times 10^{10}$ GeV and $\la
\phi \ra \simeq 1 \times 10^{16}$ GeV for $m_{3/2}=100$~TeV.  Note
that $m_{3/2} \ll 1\,$TeV cannot be realized unless $g \gg 1$. 
From Fig.~\ref{fig:m-vev}, we can see that the new inflation model is
excluded for $m_{3/2}= 1$ TeV with $B_h = 1$, while it is below the
bound for $m_{3/2} = 100$~TeV.

\subsubsection{Multiple-field inflation model}

Next we consider an inflation model with multiple fields.  Among many
multiple-field inflation models proposed so far, there is an important
class of models described by the following superpotential:
\beq
\label{eq:multi-generic}
W(\phi,\psi)\;=\; \phi f(\psi),
\eeq
where $f(\psi)$ is a function of $\psi$.  The potential minimum in the
global SUSY limit is located at
\bea
\la \phi \ra&=&0,\non\\
\la \psi \ra &=& \psi_0,
\label{eq:min-global-generic}
\eea
where $\psi_0$ satisfies $f(\psi_0)=0$.  Note that the true minimum is
slightly displaced from (\ref{eq:min-global-generic}), once the SUSY
breaking field is taken into
account~\cite{Kawasaki:2006gs,Dvali:1997uq}.

For instance, the above class of the models includes a new inflation
model~\cite{Asaka:1999jb} and a hybrid inflation model~\cite{Copeland:1994vg}, 
described by
\beq
\label{eq:multi}
W(\phi,\psi)\;=\; \phi \left(\mu^2 - \frac{\psi^n}{M^{n-2}} \right),
\eeq
where $\mu$ determines the inflation energy scale and $M$ is an
effective cut-off scale. In the new inflation model $\psi$ plays a
role of the inflaton, while $\phi$ is the inflaton in the hybrid
inflation model.

The inflaton fields $\phi$ and $\psi$ have almost the same masses,
\beq
\label{eq:susy-mass-2field}
m_\phi \;\simeq\; m_\psi \;\simeq\; \left| e^{G/2} \nabla_\phi G_\psi \right|,  
\eeq
which are assumed to be much larger than the gravitino mass.  It
should be noted that $\phi$ and $\psi^\dag$ (and/or $\psi$) almost
maximally mix with each other to form the mass eigenstates due to the
almost degenerate masses~\cite{Kawasaki:2006gs}.  To see this one
should note that the difference between the diagonal components of the
mass matrix is small: $|M^2_{\phi \bar \phi}-M^2_{\psi \bar \psi}| =
O(m_{3/2}^2)$, while the off-diagonal component is relatively large:
$M^2_{\phi \psi} = O(m_{3/2} m_\phi )$, resulting in the almost
maximal mixing between $\phi$ and $\psi^\dag$.  Similar mixing may
occur between $\phi$ and $\psi$.  This mixing is effective at the
inflaton decay, since the Hubble parameter at the decay should be
smaller than $O(m_{3/2})$ to satisfy the bounds from the thermally
produced gravitinos.  The mass eigenstates are obtained after taking
account of the (almost) maximal mixing between $\phi$ and $\psi
(\psi^\dag)$:
\beq
\varphi_{\pm} \;\simeq \frac{\phi \pm \psi^{(\dag)}}{\sqrt{2}}.
\eeq
The mass-eigenstates have the mass given by (\ref{eq:susy-mass-2field})
and the effective VEV $\la \varphi_{\pm} \ra$ given by $\psi_0/\sqrt{2}$ unless there 
is cancellation.

\vspace{10mm}
{\it 2-A.~~~~ New inflation model\\ \\}
The new inflation discussed above is also realized
for~\cite{Asaka:1999jb}
\bea
K  & = & |\phi|^2 + |\psi|^2 + \frac{k_1}{4}|\phi|^4
+ k_2 |\phi|^2 |\psi|^2 + \frac{k_3}{4}|\psi|^4,\non\\
W & = & \phi (v^2 -g \,\psi^4),
\eea
in which the inflaton is $\psi$, while $\phi$ stays at the origin
during and after inflation. If one defines $k \equiv k_2 -1$, the
scalar potential for the inflaton $\psi$ becomes the same as the
single-field new inflation model, although the gravitino mass is not
related to the inflaton parameters.  After the inflation ends, the
energy of the universe is dominated by the oscillation energy of
$\psi$. Although $\phi_0$ is suppressed compared to $\psi_0$, the
effective VEV is given by $\psi_0 /\sqrt{2}$, since $\phi$ and
$\psi^\dag$ almost maximally mixes with each other in the vacuum.
Thus the constraint on this model is comparable to that on the
single-field new inflation.  We plot the values of $m_\phi$ and $\la
\varphi_\pm \ra$ for $g =10^{-4} - 1$ and $k=10^{-4}-10^{-1.5}$ with
the e-folding number $N=50$ in Figs.~\ref{fig:contour} and
\ref{fig:m-vev}.  The (multi-field) new inflation model is excluded
for $m_{3/2} = 1$\,TeV with $B_h = 1$, while it is allowed for
$m_{3/2} = 1 {\rm\,GeV}$ and $100 {\rm\, TeV}$.

\vspace{10mm}
{\it 2-B.~~~~ Hybrid  and Smooth hybrid inflation models\\ \\}
The hybrid inflation model contains two kinds of superfields: one is
$\phi$ which plays a role of inflaton and the others are waterfall
fields $\psi$ and
$\tilde{\psi}$~\cite{Copeland:1994vg}.
After inflation ends, $\phi$ as well as $\psi$($\tilde{\psi}$)
oscillates around the potential minimum and dominates the universe
until the reheating.

The superpotential $W(\phi, \psi,\tilde{\psi})$ for the inflaton
sector is
\begin{equation}
   \label{eq:spot_hyb}
   W(\phi, \psi,\tilde{\psi}) = \phi (\mu^{2}  
   - \lambda \tilde{\psi}\psi),
\end{equation}
where $\psi$ and $\tilde \psi$ are assumed to be charged under $U(1)$
gauge symmetry.  Here $\lambda$ is a coupling constant and $\mu$ is
the inflation energy scale. The potential minimum is located at
$\la\phi\ra = 0$ and $\langle \psi\rangle = \langle\tilde{\psi}\rangle
= \mu/\sqrt{\lambda}$ in the SUSY limit.  For a successful inflation,
$\mu$ and $\lambda$ are related as $\mu \simeq 2\times
10^{-3}\lambda^{1/2}$ for $\lambda \gtrsim 10^{-3}$, and $\mu \simeq
2\times 10^{-2}\lambda^{5/6}$ for $\lambda \lesssim 10^{-3}$.

Due to the D-term potential one linear combination of $\psi$ and
$\tilde \psi$, given by $\psi^{(-)} \equiv
(\psi-\tilde{\psi})/\sqrt{2}$, has a large mass of $\sim g \la \psi
\ra$ ($g$ denotes the gauge coupling), while the other, $\psi^{(+)}
\equiv (\psi+\tilde{\psi})/\sqrt{2}$ has a mass equal to that of
$\phi$: $m_{\psi^{(+)}}=m_\phi = \sqrt{2} \lambda \langle \psi
\rangle$.  It is the latter that (almost) maximally mixes with $\phi$
to form mass eigenstates.  Note that VEV of $\psi^{(+)}$ is equal to
$\sqrt{2}\langle \psi \rangle$.

For $\lambda \sim 10^{-1} -10^{-5}$~\cite{BasteroGil:2006cm} we obtain
$\mu \sim 8\times 10^{-4} - 1\times10^{-6}$, $m_{\phi} \sim 10^{15} -
10^{10} $~GeV, and $\la \varphi_\pm \ra =\mu/\sqrt{\lambda} \sim
O(10^{15})$GeV .  From Fig.~\ref{fig:m-vev}, one can see the hybrid
inflation model is excluded by the gravitino overproduction for
$m_{3/2} = 1{\rm\,TeV}$ with $B_{3/2} = 1$.  For $m_{3/2} =
1{\rm\,GeV}$ and $100\,$TeV, the constraints become slightly mild, but
a certain fraction of the parameter space is still excluded.  The
allowed parameter space corresponds to $\lambda \lesssim 10^{-2}$.
Note that the parameter space allowed by the gravitino production
leads to almost scale-invariant power spectrum, which is disfavored by
the WMAP data~\cite{Spergel:2006hy}.  It is possible to make the
scalar spectral index $n_s$ smaller than $1$ by introducing
non-renormalizable interactions in the K\"ahler
potential~\cite{BasteroGil:2006cm,ur Rehman:2006hu}.

Here we comment on interesting observation concerning the spectral
index and the cosmic string. In this type of hybrid inflation, cosmic
strings are formed after inflation because $\psi$ and $\tilde\psi$
have $U(1)$ gauge charges. As is well known, the cosmic strings
contribute to the density fluctuations.  Including the effects of the
cosmic string makes the spectral index $n_s$ between $0.98$ and $1$
compatible with the WMAP data~\cite{Battye:2006pk, Bevis:2007gh}, if
the tension of the cosmic string is $G \mu = O(10^{-7})$. According to
Ref.~\cite{Battye:2006pk}, this corresponds to the region with
$\lambda \sim O(10^{-3} - 10^{-2})$. Interestingly enough, the region
is just below the constraints from the gravitino production in the
case of $m_{3/2} = 1{\rm\,GeV}$ and $100\,$TeV~\footnote{Including the
soft terms, the inflaton dynamics is somewhat modified, and
correspondingly the inflaton parameters are slightly changed,
especially if the gravitino mass is heavy~\cite{Senoguz:2004vu}.}.
This means that, for that region, the gravitino non-thermally produced
by the inflaton decay may account for the present DM
abundance~\cite{Takahashi:2007tz}, if the gravitino is stable. For the
unstable gravitino of a mass $m_{3/2} \gtrsim O(10)\,$TeV, Wino-like
LSP produced by the gravitino decay may be the dominant component of
DM. Moreover, since the required tension of the cosmic string is
relatively large and is close to the present observational upper
bound, one may be able to discover the cosmic string in the future
observations. Since the inflaton mass and the VEV are small, it is
difficult to realize the non-thermal leptogenesis via the spontaneous
decay (see Eq.~(\ref{eq:2-body})). However, one can naturally
incorporate the non-thermal leptogenesis into the hybrid inflation
model by identifying the $U(1)$ symmetry with a $U(1)_{B-L}$
symmetry~\cite{Asaka:1999yd}.

Next let us consider a smooth hybrid inflation
model~\cite{Lazarides:1995vr}, which predicts the scalar spectral
index as $n_s \simeq 0.97$, which is slightly smaller than the simple
hybrid inflation model.  The superpotential of the inflaton sector is
\beq
 \label{eq:spot_smhyb}
   W(\phi, \psi,\tilde{\psi}) = \phi \left(\mu^{2}  
   - \frac{ (\tilde{\psi}\psi)^n}{M^{2n-2}}\right).
\eeq
The VEVs of $\psi$ and $\tilde \psi$ are given by $\la \psi \ra = \la
\tilde \psi \ra=(\mu M^{n-1})^{1/n}$, and we assume that $\psi =
\tilde{\psi}$ always holds due to the additional D-term potential.
Then one of the combination, $\psi^{(+)}\equiv
(\psi+\tilde{\psi})/\sqrt{2}$, almost maximally mixes with $\phi$ to
form the mass eigenstate of a mass $m_\phi = \sqrt{2}n \mu^2/\la \psi
\ra$.  For $n=2$ we obtain $\mu \sim 4\times 10^{-4} -9\times10^{-5}$,
and $m_{\phi} \sim 1\times 10^{14} - 6 \times 10^{14} $~GeV.  From
Fig.~\ref{fig:m-vev}, one can see that the smooth hybrid inflation
model is excluded for a broad range of $m_{3/2}$.

\vspace{10mm}
{\it 2-C.~~~~ Chaotic inflation model\\ \\}
A chaotic inflation \cite{Linde:1983gd} is realized in SUGRA, based on
a Nambu-Goldstone-like shift symmetry of the inflaton chiral multiplet
$\phi$ ~\cite{Kawasaki:2000yn}. Namely, we assume that the K\"ahler
potential $K(\phi,\phi^\dag)$ is invariant under the shift of $\phi$,
\begin{equation}
  \phi \rightarrow \phi + i\,A,
  \label{eq:shift}
\end{equation}
where $A$ is a dimensionless real parameter. Thus, the K\"ahler
potential is a function of $\phi + \phi^\dag$; $K(\phi,\phi^\dag) =
K(\phi+\phi^\dag)= c\,(\phi+\phi^\dag) + \frac{1}{2}
(\phi+\phi^\dag)^2 + \cdots$, where $c$ is a real constant and must be
smaller than $O(1)$ for a successful inflation.  As opposed to the
other inflation models, this model allows a linear term in the
K\"ahler potential.  The coefficient $c$ corresponds to the inflaton
VEV in the other models.  If there is no other symmetry such as a
$Z_2$ symmetry, there is no reason to expect that $c$ is much smaller
than unity.

We identify the imaginary part of $\phi$ with the inflaton field
$\varphi \equiv \sqrt{2} {\rm\,Im}[\phi]$.  Moreover, we introduce a
small breaking term of the shift symmetry in the superpotential in
order for the inflaton $\varphi$ to have a potential:
\begin{equation}
  W(\phi,\psi) = m\,\phi \,\psi, 
  \label{eq:mass}
\end{equation}
where we introduced a new chiral multiplet $\psi$, and $m \simeq 2
\times 10^{13}$GeV determines the inflaton mass.

 One might suspect that it is only the real component of $\phi$ that
can decay into the gravitinos, since the shift symmetry dictates that
the only real component $(\phi + \phi^\dag)$ appears in the K\"ahler
potential.  However, it is not surprising that this is not the case,
since the decay amplitude is proportional to powers of the large SUSY
mass $m$ that explicitly violates the shift symmetry.

We plot the chaotic inflation model with $c= 0.1 - 1$ in
Figs.~\ref{fig:contour} and \ref{fig:m-vev}. One can see that it is
excluded for almost entire values of $m_{3/2}$ (except for $m_{3/2}
\lesssim O(10)$\,eV).  Note however that one can avoid the constraints
by assuming an approximate $Z_2$ symmetry to suppress $c$.

\subsection{Possible solutions to the gravitino problem}
\label{sol}
Here let us briefly mention possible solutions to the gravitino
overproduction problem.  As mentioned above, one solution is to
postulate a symmetry of the inflaton. If the symmetry is unbroken at
the vacuum (or if the breaking of this symmetry is small), the VEV of
the inflaton, $\la \phi \ra$, is zero (or suppressed). As the
gravitino production rate is proportional to $\la \phi \ra ^2$, one
can avoid the gravitino overproduction for such inflation models. Note
however that, if the symmetry is exact, the visible matter fields as
well must be charged under the same symmetry, since otherwise the
inflaton cannot decay into the visible sector.  This solution can be
achieved e.g. in the chaotic inflation model; one can assign $Z_2$
symmetry on the inflaton~\cite{Endo:2006nj}. Also, there are inflation
models in which the inflaton is identified with the MSSM
fields~\cite{Connors:1988yx,Kasuya:2003iv, Allahverdi:2006iq,
Lyth:2006ec} or the right-handed sneutrino~\cite{Murayama:1992ua}.  By
similar reasoning, the gravitino overproduction from the inflaton
decay can be avoided in these models.

So far, we have assumed that there is no late-time entropy production
after the inflaton decay. If huge entropy
production~\cite{Lyth:1995ka,Kawasaki:2004rx} occurs after the
reheating of the inflaton, any pre-existing gravitinos are
diluted. However, since it also dilutes the pre-existing baryon
number, one may have to generate the baryon asymmetry after the
entropy
production~\cite{Stewart:1996ai,Kasuya:2001tp,Jeong:2004hy,Kawasaki:2006yb}.
Note also that the decay processes discussed in this paper can be
applied to any scalar fields, and so, the scalar field that induces
the large entropy may produce the gravitinos again.  One has to make
sure that this does not happen.

Another solution is to assume that the gravitino mass is either
extremely heavy or extremely light. If the gravitino mass exceeds
$O(10^6)\,$GeV, the gravitino may decay before the decoupling of the
LSP particle. To realize this, however, one has to contrive a set-up
in which anomaly-mediation is suppressed.  If we stick to the
gravitino mass smaller than $100$TeV on the basis of naturalness, the
gravitino problem sets severe bounds on the high-scale inflation
models as seen above.  On the other hand, if the gravitino mass is
lighter than $O(10)\,$eV, the gravitino is thermalized and its
cosmological abundance is negligibly small~\cite{Viel:2005qj}. So, the
gravitino problem is absent for such a very light gravitino.  Note
that, in order to ameliorate the gravitino problem instead of solving
it completely, one does not have to go to such extremes.  For the
gravitino mass moderately lighter or heavier than the weak scale, the
gravitino problem is relaxed especially for the low-scale inflation
models (see Fig.~\ref{fig:m-vev}).

If the inflaton mass is quite light, one can evade the constraints as
one can see from Fig.~\ref{fig:m-vev}.  However, it should be noted
that the reheating temperature is set to be the highest allowed value
in Fig.~\ref{fig:m-vev}. For a lower reheating temperature, the
constraints become severer. This means that, one may have to introduce
relatively strong couplings of the inflaton to the SM particles in
order to realize the reheating temperature adopted in
Fig.~\ref{fig:m-vev}.  For instance, if the inflaton has only the
dimension-five couplings to the visible sector suppressed by the
Planck scale, the reheating temperature $T_R$ is proportional to
$m_\phi^{3/2}$, which makes the gravitino abundances (\ref{Y32-pair})
and (\ref{Y32-anomaly}) rather insensitive to the inflaton mass. Then, one
cannot evade the gravitino problem simply by changing the inflaton
mass~\footnote{ This is one of the reasons why the moduli-induced
gravitino problem~\cite{moduli} is quite difficult to be solved. }.

Lastly, let us comment on the SUSY breaking sector which involves the
conformal dynamics.  As mentioned in Sec.~\ref{sec:4}, the inflaton
field with $m_\phi \gg \Lambda$ may not decay into the conformal SUSY
breaking sector. Then the gravitino production is suppressed.  This
solution is appealing because one does not have to impose non-trivial
constraints on the inflation models or on the thermal history of the
universe.  The only requisite is the conformal dynamics in the SUSY
breaking sector, which has its own phenomenological virtues
independently of the gravitino overproduction problem. Furthermore, it
may naturally lead to the successful non-thermal leptogenesis
scenario, which will be discussed elsewhere~\cite{prep}.

\subsection{Comments on preheating}
Here we would like to mention the effects of the
preheating~\cite{Kofman:1994rk}, i.e., the non-perturbative inflaton
decay process, which we have not taken into account so far. The
non-thermal gravitino production from the inflaton decay has somewhat
checkered history; it was once claimed that the gravitinos were
non-thermally produced during preheating~\cite{Kallosh:1999jj}, but it
was later concluded that the inflatino, instead of the gravitino in
the low energy, was actually created~\cite{Nilles:2001ry}. Since the
inflatino decays much earlier than the BBN
epoch~\cite{Allahverdi:2000fz}, the non-thermal `gravitino' (actually,
inflatino) production turned out to be harmless.
However, as discussed so far, we have found that the gravitinos are 
generically produced  by the perturbative decay processes.

Another concern is whether our results are modified by including the
effects of the preheating~\cite{Shuhmaher:2007pv}. We believe that our
arguments on the gravitino overproduction problem are robust and they
are not essentially modified even if the preheating occurs.  First of
all, we would like to emphasize that the decay processes discussed so
far are perturbative ones, and therefore they are always present. On
the other hand, it crucially depends both on the global structure of
the inflaton potential and on the couplings of the inflaton to matter
fields whether the preheating occurs and how efficiently it proceeds. 

Let us assume that the preheating actually occurs and it proceeds
quite efficiently without any back reaction, i.e., the inflaton
transfers most of its energy into other particles soon after the
inflaton starts oscillating. This corresponds to the instantaneous
reheating, which generically leads to the overproduction of the
gravitinos due to particle scatterings, instead of the non-thermal
production. The latter is suppressed in this case since the reheating
temperature will become high (see (\ref{Y32-pair}) and
(\ref{Y32-anomaly})). Therefore, for the most inflation models, such
an efficient preheating should not occur, since otherwise too many
gravitinos are produced by the conventional thermal scatterings.

On the other hand, if the preheating is not so efficient, then one has
to take account of the back reaction processes, and the preheating
typically ends at a certain point.  The reheating of the universe is
induced by the perturbative decay of the inflaton in the end, and the
gravitinos are generically produced by the decay.  Thus it is unlikely
that the preheating solves or ameliorates the gravitino problem; one
has to contrive a model in which the preheating proceeds quite
efficiently, but the gravitinos are not produced by the scatterings of
the decay products. Specifically, the decay products should not reach
thermal equilibrium.

\section{Conclusion}
\label{sec:6}

The present observational data on the CMB and the large-scale
structure, together with the strong theoretical motivation to resolve
severe problems in the standard big bang cosmology, have led us to
believe that the universe underwent an inflationary epoch at an early
stage.  While there are many inflation models (called as ``the
inflationary zoo''), we still do not know which inflation model is
realized in nature.  The study on the density fluctuations such as
isocurvature perturbations, non-gaussianity, tensor-mode, and their
effects on the CMB power spectrum is quite useful, but is not enough
at present to pin down the inflation model.  This is partly because of
our ignorance of thermal history of the universe beyond the standard
big bang theory, e.g., how the inflaton reheats the universe.

In this paper, we have investigated the inflaton decay processes in
the supergravity. In particular, we have shown that the gravitinos are
generically produced in the inflaton decay.  There are three different
processes for the production. One is the direct production of a pair
of the gravitinos.  This is effective especially for low-scale
inflation models. The other two are due to the inflaton decay into the
SUSY breaking sector; the spontaneous decay at the tree level and the
anomaly-induced one at the one-loop level.  Those non-thermally
produced gravitinos set tight constraints on the inflation models,
together with the constraints from thermally produced
gravitinos. Indeed, these two constraints are complementary in a sense
that the dependence on the reheating temperature is different.  For
higher $T_R$, more gravitinos are thermally produced, while the
non-thermally production becomes important for lower $T_R$.  They also
depend on the SUSY breaking scenarios. In fact, almost all parameter
space for the inflaton is excluded for the gravity-mediated SUSY
breaking scenarios.

Apart from the gravitino productions, the inflaton naturally decays
into the visible sector especially through the top Yukawa coupling and
SU(3)$_C$ gauge interaction. Thus one does not need to introduce any
direct ad hoc couplings by hand in order to induce the reheating.

The above studies may provide us with a breakthrough toward the full
understanding of the inflationary universe. In addition to the
standard analysis on the density fluctuations, the inflation models in
supergravity are subject to the constraints due to the (non)-thermally
produced gravitinos.  Whether a consistent thermal history after
inflation is realized now becomes a new guideline to sort out the
inflationary zoo, and hopefully it will pin down the true model,
together with data in the future collider experiments such as LHC.


\end{document}